\font\tenfrakturb=eufb10
\font\tenfraktur=eufm10
\font\tenmsbm=msbm10
\font\sevenfrakturb=eufb7
\font\sevenfraktur=eufm7
\font\sevenmsbm=msbm7
\font\fivefrakturb=eufb5
\font\fivefraktur=eufm5
\font\fivemsbm=msbm5
\newfam\bgothicfam
\newfam\gothicfam
\newfam\msbmfam
\textfont\bgothicfam = \tenfrakturb \scriptfont\bgothicfam=\sevenfrakturb
\scriptscriptfont\bgothicfam=\fivefrakturb
\textfont\gothicfam = \tenfraktur \scriptfont\gothicfam=\sevenfraktur
\scriptscriptfont\gothicfam=\fivefraktur
\textfont\msbmfam = \tenmsbm \scriptfont\msbmfam=\sevenmsbm
\scriptscriptfont\msbmfam=\fivemsbm

\def\Bbb{\tenmsbm\fam\msbmfam}

\catcode`@=11
\def\renewcounter#1{\@definecounter{#1}\@ifnextchar[{\@newctr{#1}}{}}

\documentstyle{elsart}
\begin{document}
\begin{frontmatter}
\title{ Estimates for parameters and characteristics of the confining 
SU(3)-gluonic field in charged pions and kaons from leptonic decays and 
chiral symmetry breaking}
\author{Yu. P. Goncharov}
\address{Theoretical Group, Experimental Physics Department, State Polytechnical
         University, Sankt-Petersburg 195251, Russia}

\begin{abstract} 
The confinement mechanism proposed earlier by the author is employed for to 
compute the decay constants $f_P$ corresponding to leptonic decays 
$P\to l^\pm+\nu_l$, $l=\mu, e$, where $P$ stands for any meson from $\pi^\pm$, 
$K^\pm$. For this aim the weak axial form factor of $P$-meson is nonperturbatively 
calculated. The study entails estimates for parameters of the confining 
SU(3)-gluonic field in charged pions and kaons. The corresponding estimates of 
the gluon concentrations, electric and magnetic colour field strengths are 
also adduced for the mentioned field at the scales of the mesons under 
consideration. Further the obtained results are applied to the problem of 
chiral symmetry breaking in quantum chromodynamics (QCD). It is shown 
that in chirally symmetric world masses of pions and kaons are fully determined 
by the confining SU(3)-gluonic field among (massless) $u$, $d$ and $s$ quarks 
and not equal to zero. Accordingly chiral symmetry is sufficiently rough 
approximate one holding true only when neglecting the mentioned SU(3)-gluonic 
field between quarks and no additional mechanism of the spontaneous chiral 
symmetry breaking connected to the so-called Goldstone bosons is required. 
Finally, a possible relation of the results obtained with a phenomenological 
string-like picture of confinement is discussed too. 
\end{abstract}
\end{frontmatter}   
\section{Introduction}

In Refs. \cite{{Gon01},{Gon051},{Gon052}} for the Dirac-Yang-Mills 
system derived from 
QCD-Lagrangian an unique family of compatible 
nonperturbative solutions was found and explored, 
which could pretend to describing confinement of two quarks. 
The applications of the family to the description of both the heavy quarkonia 
spectra \cite{{Gon03},{Gon08a}} and a number of properties of pions, kaons, 
$\eta$- and $\eta^\prime$-mesons \cite{{Gon06},{Gon07a},{Gon07b}} 
showed that the confinement 
mechanism is qualitatively the same for both light mesons and heavy quarkonia.
At this moment it can be described in the following way.

Two main physical reasons underlie linear confinement in the 
mechanism under discussion. The first one is that gluon exchange between 
quarks is realized with the propagator different from the photon-like one and 
existence and form of such a propagator is a {\em direct} consequence of the 
unique confining nonperturbative solutions of the Yang-Mills equations 
\cite{{Gon051},{Gon052}}. The second reason is that, 
owing to the structure of mentioned propagator, quarks mainly emit and 
interchange the soft gluons so the gluon condensate (a classical gluon field) 
between quarks basically consists of soft gluons (for more details 
see Refs. \cite{{Gon051},{Gon052}}) but, because of that any gluon also emits 
gluons (still softer), the corresponding gluon concentrations 
rapidly become huge and form a linear confining magnetic colour field of 
enormous strengths, which leads to the confinement of quarks. This is by virtue of 
the fact that just magnetic part of the mentioned propagator is responsible 
for a larger portion of gluon concentrations at large distances since the 
magnetic part has stronger infrared singularities than the electric one. 
Under the circumstances 
physically nonlinearity of the Yang-Mills equations effectively vanishes so the 
latter possess the unique nonperturbative confining solutions of the 
Abelian-like form (with the values in Cartan subalgebra of SU(3)-Lie algebra) 
\cite{{Gon051},{Gon052}} which describe 
the gluon condensate under consideration. Moreover, since the overwhelming majority 
of gluons is soft they cannot leave hadron (meson) until some gluons obtain  
additional energy (due to an external reason) to rush out. So we also deal with 
confinement of gluons. 

The approach under discussion equips us with the explicit wave functions 
for every two quarks (meson or quarkonium). The wave functions are parametrized 
by a set of real constants $a_j, b_j, B_j$ describing the mentioned 
{\em nonperturbative} confining SU(3)-gluonic field (the gluon condensate), and 
they are {\em nonperturbative} modulo square integrable 
solutions of the Dirac equation in the above confining SU(3)-field and also  
depend on $\mu_0$, the reduced
mass of the current masses of quarks forming meson. It is clear that under the 
given approach just constants $a_j, b_j, B_j,\mu_0$ determine all properties 
of any meson (quarkonium), i. e.,  the approach directly appeals to quark 
and gluonic degrees of freedom as should be according to the first principles 
of QCD. Also it is clear that the constants mentioned should be extracted from 
experimental data. 

Such a program has been to a certain extent advanced in 
Refs. \cite{{Gon03},{Gon08a},{Gon06},{Gon07a},{Gon07b}}. The aim of 
the present paper is to continue obtaining estimates for $a_j, b_j, B_j$ for 
concrete mesons starting from experimental data on spectroscopy of one or 
another meson. We here consider charged pions and kaons and their leptonic 
decays $P\to l^\pm+\nu_l$, $l=\mu, e$, where $P$ stands for any meson 
from $\pi^\pm$, $K^\pm$. 

Of course, when conducting our considerations 
we shall rely on the standard quark model (SQM) based on SU(3)-flavor symmetry 
(see, e. g., Ref. \cite{pdg}), so in accordance with SQM 
$\pi^\pm=u\bar{d}$, $\bar{u}d$, $K^\pm=u\bar{s}$, $\bar{u}s$ respectively. 

Under the circumstances Section 2 contains main relations underlying 
our approach. Section 3 is devoted to computing the electric form factor, the 
root-mean-square radius $<r>$ 
and the magnetic moment of the mesons under consideration in an explicit 
analytic form while Section 4 gives an explicit analytic expression of 
the weak axial form factor for those mesons. Results of Sections 3 and 4 
are used in Section 5 for obtaining estimates for parameters of the confining 
SU(3)-gluonic field for the mesons under discussion. 
Section 6 employs the obtained parameters of SU(3)-gluonic 
field to get the corresponding estimates for such characteristics of the 
mentioned field as gluon concentrations, electric and magnetic colour field 
strengths at the scales of the mesons in question. In Section 7 results of 
previous sections are applied to the problem of chiral symmetry breaking in 
QCD. Section 8 explores a possible connection of the results obtained 
with a phenomenological string-like picture of confinement while Section 9 is 
devoted to discussion and concluding remarks. 
                      
At last, Appendices A and B contain the detailed description 
of main building blocks for meson wave functions in the approach under 
discussion, respectively: eigenspinors of the Euclidean Dirac operator on 
2-sphere ${\Bbb S}^2$ and radial parts for the modulo square integrable 
solutions of Dirac equation in the confining SU(3)-Yang-Mills field. 
 
Further we shall deal with the metric of
the flat Minkowski spacetime $M$ that
we write down (using the ordinary set of local spherical coordinates
$r,\vartheta,\varphi$ for the spatial part) in the form
$$ds^2=g_{\mu\nu}dx^\mu\otimes dx^\nu\equiv
dt^2-dr^2-r^2(d\vartheta^2+\sin^2\vartheta d\varphi^2)\>. \eqno(1)$$
Besides, we have $|\delta|=|\det(g_{\mu\nu})|=(r^2\sin\vartheta)^2$
and $0\leq r<\infty$, $0\leq\vartheta<\pi$,
$0\leq\varphi<2\pi$.

Throughout the paper we employ the Heaviside-Lorentz system of units 
with $\hbar=c=1$, unless explicitly stated otherwise, so the gauge coupling 
constant $g$ and the strong coupling constant ${\alpha_s}$ are connected by 
relation $g^2/(4\pi)=\alpha_s$. 
Further, we shall denote by $L_2(F)$ the set of the modulo square integrable
complex functions on any manifold $F$ furnished with an integration measure, 
then $L^n_2(F)$ will be the $n$-fold direct product of $L_2(F)$
endowed with the obvious scalar product while $\dag$ and $\ast$ stand, 
respectively, for Hermitian and complex conjugation. Our choice of Dirac 
$\gamma$-matrices conforms to the so-called standard representation and is 
the same as in Ref. \cite{Gon06}. At last $\otimes$ means 
the tensorial product of matrices and $I_n$ is the unit $n\times n$ matrix 
so that, e.g., we have 
$$I_3\otimes\gamma^\mu=
\pmatrix{\gamma^\mu&0&0\cr 0&\gamma^\mu&0\cr 0&0&\gamma^\mu\cr}$$ 
for any Dirac $\gamma$-matrix $\gamma^\mu$ and so forth.

When calculating we apply the 
relations $1\ {\rm GeV^{-1}}\approx0.1973269679\ {\rm fm}\>$,
$1\ {\rm s^{-1}}\approx0.658211915\times10^{-24}\ {\rm GeV}\>$, 
$1\ {\rm V/m}\approx0.2309956375\times 10^{-23}\ {\rm GeV}^2$, 
$1\ {\rm T}=4\pi\times10^{-7} {\rm H/m}\times1\ {\rm A/m}
\approx0.6925075988\times 10^{-15}\ {\rm GeV}^2 $. 

Finally, for the necessary estimates we shall employ the $T_{00}$-component 
(volumetric energy density ) of the energy-momentum tensor for a 
SU(3)-Yang-Mills field which should be written in the chosen system of units 
in the form
$$T_{\mu\nu}=-F^a_{\mu\alpha}\,F^a_{\nu\beta}\,g^{\alpha\beta}+
{1\over4}F^a_{\beta\gamma}\,F^a_{\alpha\delta}g^{\alpha\beta}g^{\gamma\delta}
g_{\mu\nu}\>. \eqno(2) $$

\section{Main relations}

As was mentioned above, our considerations shall be based on the unique family 
of compatible nonperturbative solutions for 
the Dirac-Yang-Mills system (derived from QCD-Lagrangian) studied at the whole 
length in Refs. \cite{{Gon01},{Gon051},{Gon052}}.  Referring for more details 
to those references, let us briefly describe and specify only the relations 
necessary to us in the present paper. 

One part of the mentioned family is presented by the unique nonperturbative 
confining solution of the SU(3)-Yang-Mills 
equations for gluonic field $A=A_\mu dx^\mu=
A^a_\mu \lambda_adx^\mu$ ($\lambda_a$ are the 
known Gell-Mann matrices, $\mu=t,r,\vartheta,\varphi$, $a=1,...,8$) and looks 
as follows 
$$ {\cal A}_{1t}\equiv A^3_t+\frac{1}{\sqrt{3}}A^8_t =-\frac{a_1}{r}+A_1 \>,
{\cal A}_{2t}\equiv -A^3_t+\frac{1}{\sqrt{3}}A^8_t=-\frac{a_2}{r}+A_2\>,$$
$${\cal A}_{3t}\equiv-\frac{2}{\sqrt{3}}A^8_t=\frac{a_1+a_2}{r}-(A_1+A_2)\>, $$
$$ {\cal A}_{1\varphi}\equiv A^3_\varphi+\frac{1}{\sqrt{3}}A^8_\varphi=
b_1r+B_1 \>,
{\cal A}_{2\varphi}\equiv -A^3_\varphi+\frac{1}{\sqrt{3}}A^8_\varphi=
b_2r+B_2\>,$$
$${\cal A}_{3\varphi}\equiv-\frac{2}{\sqrt{3}}A^8_\varphi=
-(b_1+b_2)r-(B_1+B_2)\> \eqno(3)$$
with the real constants $a_j, A_j, b_j, B_j$ parametrizing the family. 
The word {\em unique} should be understood in the strict mathematical sense. 
In fact in Ref. \cite{Gon051} the following theorem was proved:

{\em The unique exact spherically symmetric (nonperturbative) solutions (i.e. 
depending only on $r$) of SU(3)-Yang-Mills equations in Minkowski spacetime 
consist of the family of (3)}.

It should be noted that solution (3) was found early in 
Ref. \cite{Gon01} but its uniqueness was proved just in Ref. \cite{Gon051} 
(see also Ref. \cite{Gon052}). Besides, in Ref. \cite{Gon051} (see also 
Ref. \cite{Gon06}) it was shown that the above unique confining solutions (3) 
satisfy the so-called Wilson confinement criterion \cite{Wil}. Up to now 
nobody contested this result so if we want to describe interaction between 
quarks by spherically symmetric SU(3)-fields then they can be only the ones 
from the above theorem. 

As has been repeatedly explained in 
Refs. \cite{{Gon051},{Gon052},{Gon03},{Gon06}}, parameters $A_{1,2}$ of 
solution (3) are inessential for physics in question and we can 
consider $A_1=A_2=0$. Obviously we have 
$\sum_{j=1}^{3}{\cal A}_{jt}=\sum_{j=1}^{3}{\cal A}_{j\varphi}=0$ which 
reflects the fact that for any matrix 
${\cal T}$ from SU(3)-Lie algebra we have ${\rm Tr}\,{\cal T}=0$. 
Also, as has been repeatedly discussed by us earlier (see, e. g., 
Refs. \cite{{Gon051},{Gon052}}), from the above form it is clear that 
the solution (3) is a configuration describing the electric Coulomb-like colour 
field (components $A^{3,8}_t$) and the magnetic colour field linear in $r$ 
(components $A^{3,8}_\varphi$) and we wrote down
the solution (3) in the combinations that are just 
needed further to insert into the Dirac equation (4). 

For the sake of completeness one should note that the similar unique confining 
solutions exist for all semisimple and 
non-semisimple compact Lie groups, in particular, for SU($N$) with $N\ge2$ and 
U($N$) with $N\ge1$ \cite{{Gon051},{Gon052}}. Explicit form of solutions, 
e.g., for SU($N$) with 
$N=2,4$ can be found in Ref. \cite{Gon052} but it should be emphasized that 
components linear in $r$ always represent the magnetic colour field in all the 
mentioned solutions. 

Another part of the compatible nonperturbative solutions for the SU(3)-
Dirac-Yang-Mills system is given by the {\em unique} nonperturbative modulo 
square integrable solutions of the Dirac equation in the confining 
SU(3)-field of (3) $\Psi=(\Psi_1, \Psi_2, \Psi_3)$ 
with the four-dimensional Dirac spinors 
$\Psi_j$ representing the $j$th colour component of the meson, 
so $\Psi$ may describe relative motion (relativistic bound states) of two quarks 
in mesons and the mentioned Dirac equation looks as follows 
$$i\partial_t\Psi\equiv  
i\pmatrix{\partial_t\Psi_1\cr \partial_t\Psi_2\cr \partial_t\Psi_3\cr}=
H\Psi\equiv\pmatrix{H_1&0&0\cr 0&H_2&0\cr 0&0&H_3\cr}
\pmatrix{\Psi_1\cr\Psi_2\cr\Psi_3\cr}=
\pmatrix{H_1\Psi_1\cr H_2\Psi_2\cr H_3\Psi_3\cr}
                   \,,\eqno(4)$$
where Hamiltonian $H_j$ is 
$$H_j=\gamma^0\left[\mu_0-i\gamma^1\partial_r-i\gamma^2\frac{1}{r}
\left(\partial_\vartheta+\frac{1}{2}\gamma^1\gamma^2\right)-
i\gamma^3\frac{1}{r\sin{\vartheta}}
\left(\partial_\varphi+\frac{1}{2}\sin{\vartheta}\gamma^1\gamma^3
+\frac{1}{2}\cos{\vartheta}\gamma^2\gamma^3\right)\right]$$
$$-g\gamma^0\left(\gamma^0{\cal A}_{jt}+\gamma^3\frac{1}{r\sin{\vartheta}}
{\cal A}_{j\varphi}\right) \eqno(5)  $$                           
with the gauge coupling constant $g$ while $\mu_0$ is a mass parameter and one 
should consider it to be the reduced mass which is equal, {\it e. g.}, for 
quarkonia, to half the current mass of quarks forming a quarkonium.

Then the unique nonperturbative modulo square integrable solutions of (4) 
are (with Pauli matrix $\sigma_1$, for more details see Refs. 
\cite{{Gon01},{Gon052}})  
$$\Psi_j=e^{-i\omega_j t}\psi_j\equiv 
e^{-i\omega_j t}r^{-1}\pmatrix{F_{j1}(r)\Phi_j(\vartheta,\varphi)\cr\
F_{j2}(r)\sigma_1\Phi_j(\vartheta,\varphi)}\>,j=1,2,3\eqno(6)$$
with the 2D eigenspinor $\Phi_j=\pmatrix{\Phi_{j1}\cr\Phi_{j2}}$ of the
Euclidean Dirac operator ${\cal D}_0$ on the unit sphere ${\Bbb S}^2$, while 
the coordinate $r$ stands for the distance between quarks. 
The explicit form of 
$\Phi_j$ is discussed in Appendix A. We can call the quantity $\omega_j$ 
relative energy of $j$th colour component of meson (while $\psi_j$ is wave 
function of a stationary state for $j$th colour component) but we can see that 
if we want to interpret (4) as equation for eigenvalues of the relative 
motion energy, i. e.,  to rewrite it in the form $H\psi=\omega\psi$ with 
$\psi=(\psi_1, \psi_2, \psi_3)$ then we should put $\omega=\omega_j$ for 
any $j$ so that $H_j\psi_j=\omega_j\psi_j=\omega\psi_j$. Under this situation, 
if a meson is composed of quarks $q_{1,2}$ with different flavours then 
the energy spectrum of the meson will be given 
by $\epsilon=m_{q_1}+m_{q_2}+\omega$ with the current quark masses $m_{q_k}$ (
rest energies) of the corresponding quarks. On the other hand for 
determination of $\omega_j$ the following quadratic equation can be obtained 
\cite{{Gon01},{Gon051},{Gon052}}
$$[g^2a_j^2+(n_j+\alpha_j)^2]\omega_j^2-
2(\lambda_j-gB_j)g^2a_jb_j\,\omega_j+
[(\lambda_j-gB_j)^2-(n_j+\alpha_j)^2]g^2b_j^2-
\mu_0^2(n_j+\alpha_j)^2=0\>,  \eqno(7)   $$
that yields (at $g\ne0$) 
$$\omega_j=\omega_j(n_j,l_j,\lambda_j)=$$ 
$$\frac{\Lambda_j g^2a_jb_j\pm(n_j+\alpha_j)
\sqrt{(n_j^2+2n_j\alpha_j+\Lambda_j^2)\mu_0^2+g^2b_j^2(n_j^2+2n_j\alpha_j)}}
{n_j^2+2n_j\alpha_j+\Lambda_j^2}\>, j=1,2,3\>,\eqno(8)$$

where $a_3=-(a_1+a_2)$, $b_3=-(b_1+b_2)$, $B_3=-(B_1+B_2)$, 
$\Lambda_j=\lambda_j-gB_j$, $\alpha_j=\sqrt{\Lambda_j^2-g^2a_j^2}$, 
$n_j=0,1,2,...$, while $\lambda_j=\pm(l_j+1)$ are
the eigenvalues of Euclidean Dirac operator ${\cal D}_0$ 
on unit sphere with $l_j=0,1,2,...$. It should be noted that in the  
papers \cite{{Gon01},{Gon051},{Gon052},{Gon03},{Gon06}} we used the ansatz (6) 
with the factor $e^{i\omega_j t}$ instead of $e^{-i\omega_j t}$ but then the 
Dirac equation (4) would look as $-i\partial_t\Psi= H\Psi$ and in equation (7) 
the second summand would have the plus sign while the first summand in numerator 
of (8) would have the minus sign. In the papers 
\cite{{Gon08a},{Gon07a},{Gon07b}}   
we returned to the conventional form of 
writing Dirac equation and this slightly modified the equations (7)--(8). In 
the given paper we conform to the same prescription as in 
Refs. \cite{{Gon08a},{Gon07a},{Gon07b}}. 

In line with the above we should have $\omega=\omega_1=\omega_2=\omega_3$ in 
energy spectrum $\epsilon=m_{q_1}+m_{q_2}+\omega$ for any meson (quarkonium) 
and this at once imposes two conditions on parameters $a_j,b_j,B_j$ when 
choosing some experimental value for $\epsilon$ at the given current quark 
masses $m_{q_1},m_{q_2}$. 

The general form of the radial parts of (6) is considered in Appendix B. 
Within the given paper we need only of the radial parts of (6) at $n_j=0$ 
(the ground state) that are [see (B.5)]  
$$F_{j1}=C_jP_jr^{\alpha_j}e^{-\beta_jr}\left(1-
\frac{gb_j}{\beta_j}\right), P_j=gb_j+\beta_j, $$
$$F_{j2}=iC_jQ_jr^{\alpha_j}e^{-\beta_jr}\left(1+
\frac{gb_j}{\beta_j}\right), Q_j=\mu_0-\omega_j\eqno(9)$$
with $\beta_j=\sqrt{\mu_0^2-\omega_j^2+g^2b_j^2}$ while $C_j$ is determined 
from the normalization condition
$\int_0^\infty(|F_{j1}|^2+|F_{j2}|^2)dr=\frac{1}{3}$. 
Consequently, we shall gain that $\Psi_j\in L_2^{4}({\Bbb R}^3)$ at any 
$t\in{\Bbb R}$ and, as a result,
the solutions of (6) may describe relativistic bound states (mesons) 
with the energy (mass) spectrum $\epsilon$.
\subsection{Nonrelativistic and the weak coupling limits}
It is useful to specify the nonrelativistic limit (when 
$c\to\infty$) for spectrum (8). For that one should replace 
$g\to g/\sqrt{\hbar c}$, 
$a_j\to a_j/\sqrt{\hbar c}$, $b_j\to b_j\sqrt{\hbar c}$, 
$B_j\to B_j/\sqrt{\hbar c}$ and, expanding (8) in $z=1/c$, we shall get
$$\omega_j(n_j,l_j,\lambda_j)=$$
$$\pm\mu_0c^2\left[1\mp
\frac{g^2a_j^2}{2\hbar^2(n_j+|\lambda_j|)^2}z^2\right]
+\left[\frac{\lambda_j g^2a_jb_j}{\hbar(n_j+|\lambda_j|)^2}\,
\mp\mu_0\frac{g^3B_ja_j^2f(n_j,\lambda_j)}{\hbar^3(n_j+|\lambda_j|)^{7}}\right]
z\,+O(z^2)\>,\eqno(10)$$
where 
$f(n_j,\lambda_j)=4\lambda_jn_j(n_j^2+\lambda_j^2)+
\frac{|\lambda_j|}{\lambda_j}\left(n_j^{4}+6n_j^2\lambda_j^2+\lambda_j^4
\right)$. 

As is seen from (10), at $c\to\infty$ the contribution of linear magnetic 
colour field (parameters $b_j, B_j$) to spectrum really vanishes and spectrum 
in essence becomes the purely nonrelativistic Coulomb one (modulo the rest energy). Also it is 
clear that when $n_j\to\infty$, $\omega_j\to\pm\sqrt{\mu_0^2+g^2b_j^2}$. 

At last, one should specify the weak 
coupling limit of (8), i.e., the case $g\to0$. As is not complicated to see 
from (8), $\omega_j\to\pm\mu_0$ when $g\to0$. But then quantities 
$\beta_j=\sqrt{\mu_0^2-\omega_j^2+g^2b_j^2}\to0$ and wave functions of (9) 
cease to be the modulo square integrable ones at $g=0$, i.e., they cease to 
describe relativistic bound states. Accordingly, this means that equation 
(8) does not make physical meaning at $g=0$. 

We may seemingly use (8) with various combinations of signes ($\pm$) before 
second summand in numerators of (8) but, due to (10), it is 
reasonable to take all signs equal to plus which is our choice within the 
paper. Besides, 
as is not complicated to see, radial parts in nonrelativistic limit have 
the behaviour of form $F_{j1},F_{j2}\sim r^{l_j+1}$, which allows one to call 
quantum number $l_j$ angular momentum for $j$th colour component though angular 
momentum is not conserved in the field (3) \cite{{Gon01},{Gon052}}. So for 
meson (quarkonium) under consideration we should put all $l_j=0$. 

\subsection{Eigenspinors with $\lambda=\pm1$}
Finally it should be noted that spectrum (8) is degenerated owing to 
degeneracy of eigenvalues for the
Euclidean Dirac operator ${\cal D}_0$ on the unit sphere ${\Bbb S}^2$. Namely,  
each eigenvalue of ${\cal D}_0$ $\lambda =\pm(l+1), l=0,1,2...$, has 
multiplicity $2(l+1)$ so we has $2(l+1)$ eigenspinors orthogonal to each other. 
Ad referendum we need eigenspinors corresponding to $\lambda =\pm1$ ($l=0$) 
so here is their explicit form [see $({\mathrm A}.16)$] 
$$\lambda=-1: \Phi=\frac{C}{2}\pmatrix{e^{i\frac{\vartheta}{2}}
\cr e^{-i\frac{\vartheta}{2}}\cr}e^{i\varphi/2},\> {\rm or}\>\>
\Phi=\frac{C}{2}\pmatrix{e^{i\frac{\vartheta}{2}}\cr
-e^{-i\frac{\vartheta}{2}}\cr}e^{-i\varphi/2},$$
$$\lambda=1: \Phi=\frac{C}{2}\pmatrix{e^{-i\frac{\vartheta}{2}}\cr
e^{i\frac{\vartheta}{2}}\cr}e^{i\varphi/2}, \> {\rm or}\>\>
\Phi=\frac{C}{2}\pmatrix{-e^{-i\frac{\vartheta}{2}}\cr
e^{i\frac{\vartheta}{2}}\cr}e^{-i\varphi/2} 
\eqno(11) $$
with the coefficient $C=1/\sqrt{2\pi}$ (for more details see 
Appendix A). 

\section{Electric form factor, the root-mean-square radius and magnetic moment}

Within the present paper we shall use relations (8) at $n_j=0=l_j$ so 
energy (mass) of mesons under consideration is given by 
$m_P=m_{q_1}+m_{q_2}+\omega$ with 
$\omega=\omega_j(0,0,\lambda_j)$ for any $j=1,2,3$ whereas 
$$\omega=\frac{g^2a_1b_1}{\Lambda_1}+\frac{\alpha_1\mu_0}
{|\Lambda_1|}=\frac{g^2a_2b_2}{\Lambda_2}+\frac{\alpha_2\mu_0}
{|\Lambda_2|}=\frac{g^2a_3b_3}{\Lambda_3}+\frac{\alpha_3\mu_0}
{|\Lambda_3|}=
\cases{m_{\pi^\pm}-m_u-m_d,\cr
 m_{K^\pm}-m_u-m_s,\cr}
\>\eqno(12)$$
and, as a consequence, the corresponding meson (quarkonium) wave functions of 
(6) are represented by (9) and (11). 
\subsection{Choice of quark masses and the gauge coupling constant}
It is evident for employing the above relations we have to assign some values 
to quark masses and gauge coupling constant $g$. In accordance with 
Ref. \cite{pdg}, at present the current quark masses necessary to us are 
restricted to intervals $1.5\>{\rm MeV}\le m_u\le 3\>\,{\rm MeV}$, 
$3.0\> {\rm MeV}\le m_d\le 7 \> {\rm MeV}$, 
$95\> {\rm MeV}\le m_s\le 120 \> {\rm MeV}$,
so we take $m_u=(1.5+3)/2\>\,{\rm MeV}=2.25\>\,{\rm MeV}$, 
$m_d=(3+7)/2\>\,{\rm MeV}=5\>\,{\rm MeV}$, 
$m_s=(95+120)/2\>\,{\rm MeV}=107.5\>\,{\rm MeV}$. 
Under the circumstances, the reduced mass $\mu_0$ of (5) will be 
$m_um_d/(m_u+m_d)$ or $m_um_s/(m_u+m_s)$. As to 
gauge coupling constant $g=\sqrt{4\pi\alpha_s}$, it should be noted that 
recently some attempts have been made to generalize the standard formula
for $\alpha_s=\alpha_s(Q^2)=12\pi/[(33-2n_f)\ln{(Q^2/\Lambda^2)}]$ ($n_f$ is 
number of quark flavours) holding true at the momentum transfer 
$\sqrt{Q^2}\to\infty$ 
to the whole interval $0\le \sqrt{Q^2}\le\infty$. We shall employ one such a 
generalization used in Refs. \cite{De1}. It looks as follows 
($x=\sqrt{Q^2}$ in GeV) 
$$ \alpha(x)=\frac{12\pi}{(33-2n_f)}\frac{f_1(x)}{\ln{\frac{x^2+f_2(x)}
{\Lambda^2}}} 
\eqno(13) $$
with 
$$f_1(x)=
1+\left(\left(\frac{(1+x)(33-2n_f)}{12}\ln{\frac{m^2}{\Lambda^2}}-1
\right)^{-1}+0.6x^{1.3}\right)^{-1}\>,\>f_2(x)=m^2(1+2.8x^2)^{-2}\>,$$
wherefrom one can conclude that $\alpha_s\to \pi=3.1415...$ when $x\to 0$, 
i. e., $g\to{2\pi}=6.2831...$. We used (13) at $m=1$ GeV, $\Lambda=0.234$ GeV, 
$n_f=3$, $x=m_{\pi^\pm}=139.56995$ MeV or $x=m_{K^\pm}=493.677$ MeV 
to obtain $g=6.091309951$ necessary for 
our further computations at the mass scale of $\pi^\pm$-mesons or 
$g=5.301208569$ at the mass scale of $K^\pm$-mesons.

\subsection{Electric form factor}
For each meson (quarkonium) with the wave function $\Psi=(\Psi_j)$ of (6) we 
can define 
electromagnetic current $J^\mu=\overline{\Psi}(I_3\otimes\gamma^\mu)\Psi=
(\Psi^{\dag}\Psi,\Psi^{\dag}(I_3\otimes{\bf \alpha})\Psi)=(\rho,{\bf J})$, 
${\bf \alpha}=\gamma^0{\bf\gamma}$.  
Electric form factor $f(K)$ is the Fourier transform of $\rho$
$$ f(K)= \int\Psi^{\dag}\Psi e^{-i{\bf K}{\bf r}}d^3x=\sum\limits_{j=1}^3
\int\Psi_j^{\dag}\Psi_j e^{-i{\bf K}{\bf r}}d^3x =\sum\limits_{j=1}^3f_j(K)=$$ 
$$\sum\limits_{j=1}^3
\int (|F_{j1}|^2+|F_{j2}|^2)\Phi_j^{\dag}\Phi_j
\frac{e^{-i{\bf K}{\bf r}}}{r^2}d^3x,\>
d^3x=r^2\sin{\vartheta}dr d\vartheta d\varphi\eqno(14)$$
with the momentum transfer $K$. At $n_j=0=l_j$, as is easily seen, for any  
spinor of (11) we have $\Phi_j^{\dag}\Phi_j=1/(4\pi)$, so the integrand in 
(14) does not depend on $\varphi$ and we can consider vector ${\bf K}$ to be 
directed along z-axis. Then ${\bf Kr}=Kr\cos{\vartheta}$ and with the help of 
(9) and relations (see Ref. \cite{PBM1}): $\int_0^\infty 
r^{\alpha-1}e^{-pr}dr=
\Gamma(\alpha)p^{-\alpha}$, Re $\alpha,p >0$, 
$\int_0^\infty r^{\alpha-1}e^{-pr}\pmatrix{\sin{(Kr)}\cr\cos{(Kr)}\cr}dr=
\Gamma(\alpha)(K^2+p^2)^{-\alpha/2}
\pmatrix{\sin{(\alpha\arctan{(K/p))}}\cr\cos{(\alpha\arctan{(K/p))}}\cr}$, 
Re $\alpha >-1$, 
Re $p > |{\rm Im}\, K|$, $\Gamma(\alpha+1)=\alpha\Gamma(\alpha)$, 
$\int_0^\pi e^{-iKr\cos{\vartheta}}\sin{\vartheta}d\vartheta=2\sin{(Kr)}/(Kr)$, 
we shall obtain 
$$ f(K)=\sum\limits_{j=1}^3f_j(K)=
\sum\limits_{j=1}^3\frac{(2\beta_j)^{2\alpha_j+1}}{6\alpha_j}\cdot
\frac{\sin{[2\alpha_j\arctan{(K/(2\beta_j))]}}}{K(K^2+4\beta_j^2)^{\alpha_j}}$$
$$=\sum\limits_{j=1}^3\left(\frac{1}{3}-\frac{2\alpha^2_j+3\alpha_j+1}
{6\beta_j^2}\cdot \frac{K^2}{6}\right)+O(K^4), \eqno(15)$$
wherefrom it is clear that $f(K)$ is a function of $K^2$, as should be, and 
we can determine the root-mean-square radius of meson (quarkonium) in the form 
$$<r>=\sqrt{\sum\limits_{j=1}^3\frac{2\alpha^2_j+3\alpha_j+1}
{6\beta_j^2}}.\eqno(16)$$
When calculating (15) also the fact was used that by virtue of the 
normalization condition for wave 
functions we have $C_j^2[P_j^2(1-gb_j/\beta_j)^2+Q_j^2(1+gb_j/\beta_j)^2]=
(2\beta_j)^{2\alpha_j+1}/[3\Gamma(2\alpha_j+1)]$.

On the other hand, we can directly calculate $<r>$ in accordance with the standard 
quantum mechanics rules as $<r>=\sqrt{\int r^2\Psi^{\dag}\Psi d^3x}=
\sqrt{\sum\limits_{j=1}^3\int r^2\Psi^{\dag}_j\Psi_j d^3x}$ and the 
result will be the same as in (16). So we should not call $<r>$ of (16) 
the {\em charge} radius of meson (quarkonium)-- it is just the radius of meson 
(quarkonium) determined 
by the wave functions of (6) (at $n_j=0=l_j$) with respect to strong 
interaction, i.e., radius of confinement.  
Now we should notice the expression (15) to depend on 3-vector ${\bf K}$. To 
rewrite it in the form holding true for any 4-vector $Q$, let us remind that 
according to general considerations (see, e.g., Ref. \cite{LL1}) the relation 
(15) should correspond to the so-called Breit frame where $Q^2=-K^2$ 
[when fixing metric by (1)] so it is 
not complicated to rewrite (15) for arbitrary $Q$ in the form 
$$ f(Q^2)=\sum\limits_{j=1}^3f_j(Q^2)=
\sum\limits_{j=1}^3\frac{(2\beta_j)^{2\alpha_j+1}}{6\alpha_j}\cdot
\frac{\sin{[2\alpha_j\arctan{(\sqrt{|Q^2|}/(2\beta_j))]}}}
{\sqrt{|Q^2|}(4\beta_j^2-Q^2)^{\alpha_j}}\> \eqno(17) $$
which passes on to (15) in the Breit frame. 

\subsection{Magnetic moment}
We can define the volumetric magnetic moment density by 
${\bf m}=q({\bf r}\times {\bf J})/2=q[(yJ_z-zJ_y){\bf i}+
(zJ_x-xJ_z){\bf j}+(xJ_y-yJ_x){\bf k}]/2$ with the meson charge $q$ and 
${\bf J}=\Psi^{\dag}(I_3\otimes{\bf \alpha})\Psi$. Using (6) we have in the 
explicit form 
$$J_x=\sum\limits_{j=1}^3
(F^\ast_{j1}F_{j2}+F^\ast_{j2}F_{j1})\frac{\Phi_j^{\dag}\Phi_j}
{r^2},\> 
J_y=\sum\limits_{j=1}^3
(F^\ast_{j1}F_{j2}-F^\ast_{j2}F_{j1})
\frac{\Phi_j^{\dag}\sigma_2\sigma_1\Phi_j}{r^2},\>$$
$$J_z=\sum\limits_{j=1}^3
(F^\ast_{j1}F_{j2}-F^\ast_{j2}F_{j1})
\frac{\Phi_j^{\dag}\sigma_3\sigma_1\Phi_j}{r^2}  \eqno(18)$$
with Pauli matrices $\sigma_{1,2,3}$.
Magnetic moment of meson (quarkonium) is ${\bf M}=\int_V {\bf m}d^3x$, where 
$V$ is volume 
of meson (quarkonium) (the ball of radius $<r>$). Then at $n_j=l_j=0$, as is seen from (9), 
(11), $F^\ast_{j1}=F_{j1},F^\ast_{j2}=-F_{j2}$, 
$\Phi_j^{\dag}\sigma_2\sigma_1\Phi_j=0 $ for any spinor of (11) which entails 
$J_x=J_y=0$, i.e., $m_z=0$ while $\int_V m_{x,y}d^3x=0$ because of 
turning the integral over $\varphi$ to zero, which is easy to check.
As a result, magnetic moments of mesons (quarkonia) with the 
wave functions of (6) (at $l_j=0$) are equal to zero, as should be according 
to experimental data \cite{pdg}. 

Though we can also evaluate magnetic form factor $F(Q^2)$ of meson (quarkonium) 
which is also a function of $Q^2$ (see Refs. \cite{{Gon07a},{Gon07b}}) but the 
latter will not be used in the given paper so we shall not dwell upon it. 

\section{Weak axial form factor}
\subsection{Preliminaries}
 Let us now address to leptonic decays $P\to l^\pm+\nu_l$, 
$l=\mu, e$. According to the standard theory of electroweak interaction the 
width of those decays is given by (see, e.g., Refs. \cite{{pdg},{Ok}})
$$\Gamma=\frac{G_F^2|V|^2}{8\pi}f^2_{P}m^2_lm_P
\left(1-\frac{m^2_l}{m_P^2}\right)^2
               \eqno(19)$$
with the Fermi coupling constant $G_F=1.16637\times10^{-5}$ GeV$^{-2}$, 
lepton mass $m_l$, the charged 
$P$-meson mass $m_P$, the corresponding element $V$ of the 
Cabibbo--Kobayashi--Maskawa mixing matrix ($V=V_{ud}=\cos{\vartheta_c}$ for 
$P=\pi^\pm$ 
and $V=V_{us}=\sin{\vartheta_c}$ for $P=K^\pm$ with the Cabibbo angle 
$\vartheta_c\approx13^\circ$), while the information about strong interaction 
of quarks in $P$-mesons is encoded in a decay constant $f_P$. Further 
parametrization in the form $f_P=m_P\Phi$ entails 
$\Phi\approx0.9364479961$ at $f_P=f_\pi\approx130.7$ MeV and 
$\Phi\approx0.3236934271$ at $f_P=f_K\approx159.8$ MeV \cite{pdg}. We can 
notice that the only invariant which $\Phi$ might depend on is $Q^2=m_P^2$, 
i.e. we should find 
such a function ${\cal F}(Q^2)$ for that ${\cal F}(Q^2=
m_P^2)\approx0.9364479961$ or $0.3236934271$ respectively. It is obvious from 
physical point of view that ${\cal F}$ should be connected with electroweak 
properties of $P$-mesons.

Let us now remark that, as is well known \cite{{pdg},{Ok}}, the mentioned 
leptonic decays of charged pions and kaons are governed by the axial-vector 
part of the weak charged hadronic current. We can try to construct that part 
from the $P$-meson wave function of (6) and the wave function $\Psi_0$ of vacuum 
state which may be chosen from the considerations that it should be similar 
to (6) in the form but it should not contain any explicit information about the 
parameters $a_j$, $b_j$, $B_j$, $\mu_0$ describing the $P$-meson properties in 
our approach and connected directly to quark and gluonic degrees of freedom. 
This requirement can be satisfied by $\Psi_0$ of the form 
$$\Psi_0=(\Psi_{0j})\equiv A_0\left( 
e^{-i\omega_j t}r^{-1}\pmatrix{\Phi^0_j(\vartheta,\varphi)\cr\
\sigma_1\Phi^0_j(\vartheta,\varphi)}\right)\>,j=1,2,3\eqno(20)$$
with the 2D eigenspinor $\Phi^0_j=\pmatrix{\Phi^0_{j1}\cr\Phi^0_{j2}}$ of the
Euclidean Dirac operator ${\cal D}_0$ on the unit sphere ${\Bbb S}^2$ and some 
real constant $A_0>0$ whose value will be fixed below (see Section 5) while the 
phase factor $e^{-i\omega_j t}$ is left only for convenience and has no  
influence on the subsequent calculations. 
\subsection{Weak axial form factor}
To fix spinor $\Phi^0_j$ let us consider the axial-vector part of the 
weak charged hadronic current in the form 
$A^{\mu}=\bar{\Psi}_0(I_3\otimes\gamma^\mu\gamma^5)\Psi=
(\Psi_0^{\dag}(I_3\otimes\gamma^5)\Psi,\Psi_0^{\dag}(I_3\otimes
{\bf \alpha}\gamma^5)\Psi)=(\rho_A,{\bf J_A})$, 
${\bf \alpha}=\gamma^0{\bf\gamma}$, with the wave functions $\Psi$ of (6)  
and $\Psi_0$ of (20) and compute $\rho_A$. Inasmuch as in the chosen representation 
$\gamma^5=-i\gamma^0\gamma^1\gamma^2\gamma^3=-\pmatrix{0& I_2\cr I_2 & 0\cr}$ 
then we shall have $\rho_A=-\frac{A_0}{r^2}\sum_{j=1}^{3}(F_{j2}-iF_{j1})
(\Phi^{0\dag}_j\sigma_1\Phi_j)$. Under this situation for the ground state of 
$P$-meson described by (9) and (11) it is natural to put $\Phi^{0}_j=
\sigma_1\Phi_j$ which entails $\Phi^{0\dag}_j\sigma_1\Phi_j=1/(4\pi)$ for any 
spinor of (11). Then we can define {\em the weak axial form factor} $f_A(K)$ 
for $P$-meson (by analogy with electric form factor in Section 3) as the 
Fourier transform of $\rho_A$ 
$$ f_A(K)= \int\rho_A e^{-i{\bf K}{\bf r}}d^3x =-\frac{iA_0}{4\pi}
\sum\limits_{j=1}^3C_j\int r^{\alpha_j}e^{-\beta_jr}
\left[Q_j\left(1+\frac{gb_j}{\beta_j}\right)-
P_j\left(1-\frac{gb_j}{\beta_j}\right)\right]
\frac{e^{-i{\bf K}{\bf r}}}{r^2}d^3x \>,  \eqno(21)$$
so computation along the same lines as for electric form factor of (15) 
yields the result            
$$ f_A(K)=iA_0\sum\limits_{j=1}^3 \frac{(2\beta_j)^{\alpha_j+1/2}
\Gamma(\alpha_j)}
{\sqrt{3\Gamma(2\alpha_j+1)}}\cdot\frac{{\cal P}_j-{\cal Q}_j}
{\sqrt{{\cal P}_j^2+{\cal Q}_j^2}}\cdot
\frac{\sin{[\alpha_j\arctan{(K/\beta_j)]}}}{K(K^2+\beta_j^2)^{\alpha_j/2}}=$$         
$$iA_0\sum\limits_{j=1}^3 \frac{(2\beta_j)^{\alpha_j+1/2}
\Gamma(\alpha_j)}
{\sqrt{3\Gamma(2\alpha_j+1)}}\cdot\frac{{\cal P}_j-{\cal Q}_j}
{\sqrt{{\cal P}_j^2+{\cal Q}_j^2}}\cdot
\frac{\alpha_j}{(\beta^2_j)^{\alpha_j/2}}\left(\frac{1}{\beta_j}-
\frac{\alpha_j^2+3\alpha_j+2}{6\beta_j^3}K^2+O(K^4)\right)
                              \eqno(22)$$
with ${\cal P}_j=P_j(1-{gb_j}/{\beta_j})$, 
${\cal Q}_j=O_j(1+{gb_j}/{\beta_j})$. It is clear from (22) that 
$f_A(K)$ is a function of $K^2$ and we can rewrite (22) for arbitrary 4-vector 
$Q$ as   
$$ f_A(Q^2)=iA_0\sum\limits_{j=1}^3 \frac{(2\beta_j)^{\alpha_j+1/2}
\Gamma(\alpha_j)}
{\sqrt{3\Gamma(2\alpha_j+1)}}\cdot\frac{{\cal P}_j-{\cal Q}_j}
{\sqrt{{\cal P}_j^2+{\cal Q}_j^2}}\cdot
\frac{\sin{[\alpha_j\arctan{(\sqrt{|Q^2|}/\beta_j)]}}}{\sqrt{|Q^2|}
(\beta_j^2-Q^2)^{\alpha_j/2}}
\eqno(23)$$
which passes on to (22) in the Breit frame where $Q^2=-K^2$. 

\section{Estimates for parameters of SU(3)-gluonic field in 
$P$-mesons}
\subsection{Basic equations}
Now we are able to estimate parameters $a_j, b_j, B_j$ of the confining 
SU(3)-field (3) in $P$-mesons. Though we can consider other possible 
functions of $Q^2$ connected with the mentioned above current ${\bf J}_A$ 
(on the analogy of the meson magnetic form factor, see 
Refs. \cite{{Gon07a},{Gon07b}}) but, obviously, the most natural function 
is $f_A(Q^2)$ of (23). It is reasonable, therefore, to take for the function 
${\cal F}(Q^2)$ mentioned in Subsection 4.1 just the function $|f_A(Q^2)|$, i.e., 
to put ${\cal F}(Q^2)=|f_A(Q^2)|$ so that we should have 
$|f_A(Q^2=m_P^2)|\approx0.9364479961$ or 0.3236934271 respectively for 
$P=\pi^\pm$ and $P=K^\pm$. Then, denoting the quantities 
$m_P/\beta_j=x_j$, we obtain the following equation for parameters of the 
confining SU(3)-gluonic field in $P$-mesons
$$|f_A(Q^2=m_P^2)|=A_0\left|\sum\limits_{j=1}^3 \frac{2^{\alpha_j+1/2}
\Gamma(\alpha_j)}
{\sqrt{3\beta_j\Gamma(2\alpha_j+1)}}\cdot\frac{{\cal P}_j-{\cal Q}_j}
{\sqrt{{\cal P}_j^2+{\cal Q}_j^2}}\cdot
\frac{\sin{(\alpha_j\arctan{x_j})}}{x_j(1-x_j^2)^{\alpha_j/2}}\right|
\approx\cases{0.9364479961,\cr
 0.3236934271.\cr}
\eqno(24)$$

Finally, equations (12) and (16) should also be added to (24) and the 
system obtained in such a way should be solved compatibly if taking 
$<r>\,\approx$ 0.672 fm for $P=\pi^\pm$ and $<r>\,\approx$ 0.560 fm 
for $P=K^\pm$ \cite{pdg}. While computing 
for distinctness we take all eigenvalues $\lambda_j$ ($j=$ 1, 2, 3) of the 
Euclidean Dirac operator ${\cal D}_0$ on the unit two-sphere ${\Bbb S}^2$ 
equal to 1. 
\subsection{Numerical results}
 The results of numerical compatible solving of equations (12), (16), 
and (24) are adduced in Tables 1--6 (with $\mu=m_P$).  

\begin{table}[htbp]
\caption{Gauge coupling constant, reduced mass $\mu_0$ and
parameters of the confining SU(3)-gluonic field for $\pi^\pm$-mesons}
\label{t.1}
\begin{center}
\begin{tabular}{|c|c|c|c|c|c|c|c|c|}
\hline
\small Particle & \small $ g$ & \small $\mu_0$ (\small MeV) & \small $a_1$ 
& \small $a_2$ & \small $b_1$ (\small GeV) & \small $b_2$ (\small GeV) 
& \small $B_1$ & \small $B_2$ \\
\hline
\scriptsize $\pi^\pm$---$u\bar{d}$, $\bar{u}d$  
& \scriptsize 6.09131
& \scriptsize 1.55172
& \scriptsize 0.0473002
& \scriptsize 0.0118497
& \scriptsize 0.178915 
& \scriptsize -0.119290
& \scriptsize -0.230
& \scriptsize  0.230 \\
\hline
\end{tabular}
\end{center}
\end{table}
\vskip0.5cm

\begin{table}[htbp]
\caption{Theoretical and experimental $\pi^\pm$-meson mass and radius}
\label{t.2}
\begin{center}
\begin{tabular}{|c|c|c|c|c|} 
\hline
\tiny Particle & \tiny Theoret. $\mu$ (MeV) &  \tiny Experim. $\mu$ (MeV) & 
\tiny Theoret. $<r>$ (fm)  & \tiny Experim. $<r>$ (fm)  \\
\hline
\scriptsize $\pi^\pm$---$u\bar{d}$, $\bar{u}d$   & \scriptsize $\mu= m_u+m_d+
\omega_j(0,0,1)= 139.570$ & \scriptsize 139.56995 & \scriptsize 0.673837 & 
\scriptsize 0.672 \\
\hline
\end{tabular}
\end{center}
\end{table}
\vskip0.5cm

\begin{table}[htbp]
\caption{Theoretical and experimental values of the charged pion weak axial 
form factor and constant $A_0$}
\label{t.3}
\begin{center}
\begin{tabular}{|c|c|c|c|} 
\hline
\tiny Particle &
\tiny Theoret. $|f_A(Q^2=\mu^2)|$ & \tiny Experim. $|f_A(Q^2=\mu^2)|$ 
& \tiny $A_0$ (MeV$^{1/2}$)\\ 
\hline
\scriptsize $\pi^\pm$---$u\bar{d}$, $\bar{u}d$  & \scriptsize 0.9364898401  & 
\scriptsize 0.9364479961 & \scriptsize 12.72\\ 
\hline
\end{tabular}
\end{center}
\end{table}

\begin{table}[htbp]
\caption{Gauge coupling constant, reduced mass $\mu_0$ and
parameters of the confining SU(3)-gluonic field for $K^\pm$-mesons}
\label{t.4}
\begin{center}
\begin{tabular}{|c|c|c|c|c|c|c|c|c|}
\hline
\small Particle & \small $ g$ & \small $\mu_0$ (\small MeV) & \small $a_1$ 
& \small $a_2$ & \small $b_1$ (\small GeV) & \small $b_2$ (\small GeV) 
& \small $B_1$ & \small $B_2$ \\
\hline
\scriptsize $K^\pm$---$u\bar{s}$, $\bar{u}s$  
& \scriptsize 5.30121
& \scriptsize 2.20387
& \scriptsize 0.167182
& \scriptsize -0.0557501
& \scriptsize 0.120150
& \scriptsize 0.131046
& \scriptsize -0.900
& \scriptsize  0.290 \\
\hline
\end{tabular}
\end{center}
\end{table}
\vskip0.5cm

\begin{table}[htbp]
\caption{Theoretical and experimental $K^\pm$-meson mass and radius}
\label{t.5}
\begin{center}
\begin{tabular}{|c|c|c|c|c|} 
\hline
\tiny Particle & \tiny Theoret. $\mu$ (MeV) &  \tiny Experim. $\mu$ (MeV) & 
\tiny Theoret. $<r>$ (fm)  & \tiny Experim. $<r>$ (fm)  \\
\hline
\scriptsize $K^\pm$---$u\bar{s}$, $\bar{u}s$   & \scriptsize $\mu= m_u+m_s+
\omega_j(0,0,1)= 493.677$ & \scriptsize 493.677 & \scriptsize 0.544342 & 
\scriptsize 0.560 \\
\hline
\end{tabular}
\end{center}
\end{table}
\vskip0.5cm

\begin{table}[htbp]
\caption{Theoretical and experimental values of the charged kaon weak axial 
form factor and constant $A_0$}
\label{t.6}
\begin{center}
\begin{tabular}{|c|c|c|c|} 
\hline
\tiny Particle &
\tiny Theoret. $|f_A(Q^2=\mu^2)|$ & \tiny Experim. $|f_A(Q^2=\mu^2)|$ 
& \tiny $A_0$ (MeV$^{1/2}$)\\ 
\hline
\scriptsize $K^\pm$---$u\bar{s}$, $\bar{u}s$  & \scriptsize 0.3232652803  & 
\scriptsize 0.3236934271 & \scriptsize 2.88\\ 
\hline
\end{tabular}
\end{center}
\end{table}

\section{Estimates of gluon concentrations, electric and magnetic colour field 
strengths}
Now let us remind that, according to Refs. \cite{{Gon052},{Gon06}}, one can 
confront the field (3) with $T_{00}$-component (volumetric energy 
density of the SU(3)-gluonic field) of the energy-momentum tensor (2) so that 
$$T_{00}\equiv T_{tt}=\frac{E^2+H^2}{2}=\frac{1}{2}\left(\frac{a_1^2+
a_1a_2+a_2^2}{r^4}+\frac{b_1^2+b_1b_2+b_2^2}{r^2\sin^2{\vartheta}}\right)
\equiv\frac{{\cal A}}{r^4}+
\frac{{\cal B}}{r^2\sin^2{\vartheta}}\>\eqno(25)$$
with electric $E$ and magnetic $H$ colour field strengths and with 
real ${\cal A}>0$, ${\cal B}>0$. One can also introduce magnetic colour 
induction $B=(4\pi\times10^{-7} {\rm H/m})\,H$, where $H$ in A/m.  

To estimate the gluon concentrations
we can employ (25) and, taking the quantity
$\omega= \mit\Gamma$, the full decay width of a meson, for 
the characteristic frequency of gluons we obtain
the sought characteristic concentration $n$ in the form
$$n=\frac{T_{00}}{\mit\Gamma}\> \eqno(26)$$
so we can rewrite (25) in the form 
$T_{00}=T_{00}^{\rm coul}+T_{00}^{\rm lin}$ conforming to the contributions 
from the Coulomb and linear parts of the
solution (3). This entails the corresponding split of $n$ from (26) as 
$n=n_{\rm coul} + n_{\rm lin}$. 

The parameters of Tables 1 and 4 were employed when computing and for simplicity 
we put $\sin{\vartheta}=1$ in (25). Also there were used the following 
present-day full decay widths -- for $\pi^\pm$-mesons: ${\mit\Gamma}=1/\tau$ with 
the life time $\tau=2.6033\times10^{-8}$ s and for $K^\pm$-mesons: 
${\mit\Gamma}=1/\tau$ with the life time $\tau=1.2386\times10^{-8}$ s, whereas 
the Bohr radius $a_0=0.529177249\cdot10^{5}\ {\rm fm}$ \cite{pdg}. 

Tables 7--8 contain the numerical results for $n_{\rm coul}$, $n_{\rm lin}$, 
$n$, $E$, $H$, $B$ for the mesons under discussion.
\begin{table}[htbp]
\caption{Gluon concentrations, electric and magnetic colour field strengths in 
$\pi^\pm$-mesons}
\label{t.7}
\begin{center}
\begin{tabular}{|lllllll|}
\hline
\scriptsize $\pi^\pm$---$u\bar{d}$, $\bar{u}d$: & \scriptsize 
$r_0=<r>= 0.673837 \ {\rm fm}$ & & &  & & \\
\hline 
\tiny $r$ (fm)& \tiny $n_{\rm coul}$ $ ({\rm m}^{-3}) $ & \tiny $n_{\rm lin}$ 
$ ({\rm m}^{-3}) $& \tiny $n$ (${\rm m}^{-3}) $ & \tiny $E$ $({\rm V/m})$ 
& \tiny $H$ $({\rm A/m})$ & \tiny $B$ $({\rm T})$\\
\hline
\tiny $0.1r_0$ 
& \tiny $ 0.366804\times10^{66}$   
& \tiny $ 0.141131\times10^{65}$ 
& \tiny $ 0.380917\times10^{66}$ 
& \tiny $ 0.201234\times10^{24}$  
& \tiny $ 0.530981\times10^{21}$ 
& \tiny $ 0.667250\times10^{15}$ \\
\hline
\tiny$r_0$ 
& \tiny$ 0.366804\times10^{62}$ 
& \tiny$ 0.141131\times10^{63}$ 
& \tiny$ 0.177811\times10^{63}$ 
& \tiny$ 0.201234\times10^{22}$  
& \tiny$ 0.530981\times10^{20}$  
& \tiny$ 0.667250\times10^{14}$ \\
\hline
\tiny$1.0$ 
& \tiny$ 0.756229\times10^{61}$  
& \tiny$ 0.640814\times10^{62}$ 
& \tiny$ 0.716437\times10^{62}$ 
& \tiny$ 0.913716\times10^{21}$  
& \tiny$ 0.357794\times10^{20}$  
& \tiny$ 0.449618\times10^{14}$ \\
\hline
\tiny$10r_0$ 
& \tiny$0.366804\times10^{58}$  
& \tiny$0.141131\times10^{61}$ 
& \tiny$0.141498\times10^{61}$ 
& \tiny$0.201234\times10^{20}$  
& \tiny$0.530981\times10^{19}$ 
& \tiny$0.667250\times10^{13}$ \\
\hline
\tiny$a_0$ 
& \tiny$0.964380\times10^{42}$  
& \tiny$0.228839\times10^{53}$ 
& \tiny$0.228839\times10^{53}$ 
& \tiny$0.326294\times10^{12}$ 
& \tiny$0.676133\times10^{15}$  
& \tiny$0.849654\times10^{9}$ \\
\hline
\end{tabular}
\end{center}
\end{table}

\begin{table}[htbp]
\caption{Gluon concentrations, electric and magnetic colour field strengths in 
$K^\pm$-mesons}
\label{t.8}
\begin{center}
\begin{tabular}{|lllllll|}
\hline
\scriptsize $K^\pm$---$u\bar{s}$, $\bar{u}s$: & \scriptsize 
$r_0=<r>= 0.544342 \ {\rm fm}$ & & &  & & \\
\hline 
\tiny $r$ (fm)& \tiny $n_{\rm coul}$ $ ({\rm m}^{-3}) $ & \tiny $n_{\rm lin}$ 
$ ({\rm m}^{-3}) $& \tiny $n$ (${\rm m}^{-3}) $ & \tiny $E$ $({\rm V/m})$ 
& \tiny $H$ $({\rm A/m})$ & \tiny $B$ $({\rm T})$\\
\hline
\tiny $0.1r_0$ 
& \tiny $ 0.303178\times10^{67}$   
& \tiny $ 0.195700\times10^{65}$ 
& \tiny $ 0.305135\times10^{67}$ 
& \tiny $ 0.838745\times10^{24}$  
& \tiny $ 0.906483\times10^{21}$ 
& \tiny $ 0.113912\times10^{16}$ \\
\hline
\tiny$r_0$ 
& \tiny$ 0.303178\times10^{63}$ 
& \tiny$ 0.195700\times10^{63}$ 
& \tiny$ 0.498878\times10^{63}$ 
& \tiny$ 0.838745\times10^{22}$  
& \tiny$ 0.906483\times10^{20}$  
& \tiny$ 0.113912\times10^{15}$ \\
\hline
\tiny$1.0$ 
& \tiny$ 0.266186\times10^{62}$  
& \tiny$ 0.579876\times10^{62}$ 
& \tiny$ 0.846062\times10^{62}$ 
& \tiny$ 0.248527\times10^{22}$  
& \tiny$ 0.493437\times10^{20}$  
& \tiny$ 0.620071\times10^{14}$ \\
\hline
\tiny$10r_0$ 
& \tiny$0.303178\times10^{59}$  
& \tiny$0.195700\times10^{61}$ 
& \tiny$0.198732\times10^{61}$ 
& \tiny$0.838745\times10^{20}$  
& \tiny$0.906483\times10^{19}$ 
& \tiny$0.113912\times10^{14}$ \\
\hline
\tiny$a_0$ 
& \tiny$0.339454\times10^{43}$  
& \tiny$0.207077\times10^{53}$ 
& \tiny$0.207077\times10^{53}$ 
& \tiny$0.887506\times10^{12}$ 
& \tiny$0.932460\times10^{15}$  
& \tiny$0.117176\times10^{10}$ \\
\hline
\end{tabular}
\end{center}
\end{table}

\section{Chiral symmetry breaking}
\subsection{Preliminaries}
As is known, in the late sixties of XX century, in light meson physics there 
arose notion of chiral symmetry and so far the latter has been actively 
exploited in phenomenology (see, e.g. reviews 
\cite{GL82}). In its turn, to provide real world with 
chiral symmetry breaking there was supposed that chiral symmetry had been 
spontaneously broken. On the other hand, historically in fact simultaneously 
with notion of chiral symmetry there was created standard model (SM) of 
electroweak interactions which later quarks were also included in. We should 
note that SM (with one Higgs doublet) contains some 
description of chiral symmetry breaking: current masses of quarks 
(as well as lepton masses) are acquired through Higgs mechanism so for quark 
masses $m_q$ we obtain (without taking mixings into account) 
$m_q=f_{q}v/\sqrt{2}$, where vacuum Higgs condensate $v\approx246$ GeV. But 
little is known about coupling constants $f_{q}$ and much may be elucidated 
only with discovering Higgs bosons. 

As a result, 
we at present have {\em two} postulated mechanisms for spontaneous breaking of 
(global) chiral symmetry: one is associated with SM and another one should be 
related with the so-called Goldstone bosons accompanying violation of any 
global symmetry. It is clear that both mechanisms should be connected in one 
or another way. As far as is known to us, up to now there has been not 
generally accepted (if any) recipe for reconciliation between mentioned 
mechanisms.

After QCD was created in the early seventies of XX century there appeared a 
hope that chiral symmetry breaking should be explained within framework of QCD 
and it should closely be related to confinement mechanism (see, e.g., Refs. 
\cite{pi0}). Under the situation we can remark that our approach has a 
well-defined chiral limit with perfectly clear physical meaning and in view of 
this it contains some description of chiral symmetry breaking. Let us look 
into it in more detail.
\subsection{Pion and kaon masses and chiral limit}
Let us more in detail write out expressions for pion and kaon masses from (12)
$$m_{\pi^\pm}=m_u+m_d+\frac{g^2a_1b_1}{1-gB_1}+\mu_0\frac{\sqrt{(1-gB_1)^2-g^2a_1^2}}
{|1-gB_1|}= $$
$$m_u+m_d+\frac{g^2a_2b_2}{1-gB_2}+\mu_0\frac{\sqrt{(1-gB_2)^2-g^2a_2^2}}
{|1-gB_2|}=$$
$$=m_u+m_d+\frac{g^2(a_1+a_2)(b_1+b_2)}{1+g(B_1+B_2)}+
\mu_0\frac{\sqrt{[1+g(B_1+B_2)]^2-g^2(a_1+a_2)^2}}{|1+g(B_1+B_2)|}
\>,$$
$$ \mu_0=\frac{m_um_d}{m_u+m_d},\eqno(27)$$
$$m_{K^\pm}=m_u+m_s+\frac{g^2a_1b_1}{1-gB_1}+\mu_0\frac{\sqrt{(1-gB_1)^2-g^2a_1^2}}
{|1-gB_1|}= $$
$$ m_u+m_s+\frac{g^2a_2b_2}{1-gB_2}+\mu_0\frac{\sqrt{(1-gB_2)^2-g^2a_2^2}}
{|1-gB_2|}=$$
$$=m_u+m_s+\frac{g^2(a_1+a_2)(b_1+b_2)}{1+g(B_1+B_2)}+
\mu_0\frac{\sqrt{[1+g(B_1+B_2)]^2-g^2(a_1+a_2)^2}}{|1+g(B_1+B_2)|}
\>,$$
$$ \mu_0=\frac{m_um_s}{m_u+m_s}\>.\eqno(28)$$
In chiral limit $m_{u}, m_{d}, m_{s}\to0$ 
we obtain 
$$(m_{\pi^\pm})_{chiral}\approx\frac{g^2a_1b_1}{1-gB_1}\approx
\frac{g^2a_2b_2}{1-gB_2}\approx
\frac{g^2(a_1+a_2)(b_1+b_2)}{1+g(B_1+B_2)}\approx130.8\, {\rm MeV}\ne0
\>,\eqno(29)$$
$$(m_{K^\pm})_{chiral}\approx\frac{g^2a_1b_1}{1-gB_1}\approx
\frac{g^2a_2b_2}{1-gB_2}\approx
\frac{g^2(a_1+a_2)(b_1+b_2)}{1+g(B_1+B_2)}\approx382\, {\rm MeV}\ne0
\>\eqno(30)$$
using the parameters $g, a_j, b_j, B_j$ adduced in Tables 1 and 4 respectively. 

We can see that in 
chiral limit the pion and kaon masses are completely determined only 
by the parameters $a_j, b_j, B_j$ of SU(3)-gluonic field between quarks, i.e. 
by interaction between quarks, and those masses have purely gluonic nature ! 
As a consequence, if neglecting gluon field we exactly obtain 
$(m_P)_{chiral}=0$. 
Analogously, the root-mean-square radii of pions (kaons) of (16) are  
well-defined in chiral limit and $<r>_{chiral}\approx0.673069$ fm or 
0.543223 fm with parameters $a_j, b_j, B_j$ of Tables 1 and 4 respectively and 
those values only slightly differ 
from 0.673837 fm or 0.544342 fm of Tables 2 and 5 accordingly at 
$m_u\,, m_d\,, m_s\,\ne0$. The same holds true for decay 
constants of (19) for leptonic decays $f_P=m_P|f_A(Q^2=m_P^2)|$ 
with $|f_A(Q^2=m_P^2)|$ of (24). So, even in chirally symmetric world the 
pions and kaons would have nonzero masses, the root-mean-square radii and 
decay constants $f_P$ for leptonic decays and all of those quantities would be 
determined only by SU(3)-gluonic interaction between massless quarks, i.e. 
they would have a purely gluonic nature. It should be emphasized that we 
can neglect neither Coulomb electric colour field of solution (3) 
(parameters $a_1$, $a_2$) nor magnetic colour field (parameters $b_1$, $b_2$) 
or else effect does vanish 
according to (29)--(30), i.e. both parts of SU(3)-gluonic field of (3) 
are important for confinement and mass generation in chiral limit. Moreover, 
since gluons are verily relativistic 
particles then the most part of mass for light mesons is conditioned by 
relativistic effects, as is seen from (29) and (30). 
\subsection{Quark condensate and chiral perturbation expansion}
Thus, our approach neatly says that chiral 
symmetry is rather rough approximation holding true more or less only when 
neglecting SU(3)-gluonic field between quarks. But we historically know that 
notion of chiral symmetry arose when no indications to existence of gluons 
were. So this was a reasonable approach for its time. However, after 
discovering gluons in the late seventies of XX century, main constructions 
developed within chiral symmetry approach have not been essentially changed 
to take into account the fact of existence of gluons and such a situation 
is continuing up to now (see, e.g. reviews \cite{GL82}).

Let us consider, for example, how one should interpret one of the key relations of 
chiral symmetry approach - the Gell-Mann-Oakes-Renner relation 
connecting the so-called quark condensate 
$<0|\bar{u}u+\bar{d}d|0>\approx-2(240\,{\rm MeV})^3$ (an order parameter 
characterizing chiral symmetry breaking) with $f_\pi$, $m_u,\,m_d,\, m_\pi$ 
(for more details see Refs. \cite{GL82})
$$f_\pi^2m_\pi^2=-(m_u+m_d)<0|\bar{u}u+\bar{d}d|0>\>.\eqno(31)\>$$
Then, under usual interpretation, since $f_\pi\ne0$ (experimental fact !) 
in chiral limit one should conclude that 
$m_\pi\sim0$ and identify pion with a Goldstone boson. It is clear that 
absence of gluons was tacitly supposed because when deriving (31) no gluons 
were known about.

As has been said above, however, presence of gluons between quarks entails 
that in chiral limit the left-hand side of (31) is well-defined and not 
equal to 0. Under the circumstances
$$<0|\bar{u}u+\bar{d}d|0>=-\frac{(f_\pi^2m_\pi^2)_{chiral}}{m_u+m_d}\to-\infty
\>,\eqno(32)$$
i.e., $<0|\bar{u}u+\bar{d}d|0>$ becomes unphysical, unobserved parameter. 
Similar analogy: when the speed of light $c\to\infty$ relativistic mechanics 
passes on to Newtonian one where $c$ is unobservable and no relations 
of Newtonian mechanics depend on $c$. Accordingly, the quark condensate is 
rather crude effective parameter that exists only when neglecting the fact of 
existence for gluons. We 
can, of course, try to amend (31) by changing 
$m_u+m_d\to m_u+m_d+(m_\pi)_{chiral}$ 
with $(m_\pi)_{chiral}$ of (29) but all the same $m_\pi$ will be $\sim0$ 
only if parameters of SU(3)-gluonic field $a_j,\,b_j$ are very small. 

Also we can make a comment on the so-called chiral perturbation theory (
for more details see reviews \cite{GL82}). Within the latter approach, for 
example, mass square of $P$-meson ($P=\pi^\pm,\,K^\pm$) is sought in the 
form of chiral perturbation expansion 
$$ m_P^2=xA_P+yB_P+ O(x^2,y^2)\>\eqno(33)\>$$
with current masses $x,y$ of quarks composing $P$-meson while coefficients 
$A_P,B_P$ are calculated in accordance with the rather hazy rules 
\cite{GL82}. We can, however, see from (27)--(28) that actual form of $m_P^2$ 
is 
$$ m_P^2=\left(A+x+y+ \frac{xy}{x+y}B\right)^2\>=f(x,y),\eqno(34)\>$$
where $A,B$ depend only on gluonic field. But then standard differential 
calculus says to us that function $f(x,y)$ (though being continuous at 
$x=y=0$) is not differentiable at $x=y=0$ and, accordingly, $f(x,y)$ does 
not possess the Taylor expansion of the form (33) at the point $x=y=0$. As a 
result, the expansion (33) is incorrect.  

We should, however, be somewhat careful: from nowhere it follows that 
parameters $a_j, b_j, B_j$ of the confining SU(3)-gluonic field in chirally 
symmetric world for pions and kaons 
should be the same as ones at present -- the latter were evaluated in 
accordance with (12) at nonzero $m_u$, $m_d$, $m_s$. Under the situation we 
can at least describe two scenarios of chiral symmetry breaking.

\subsection{First scenario}
In chirally symmetric world (e.g., in early stages of universe evolution) 
massless quarks interchange with gluons which generate nonzero masses of 
hadrons of purely gluonic nature, as was discussed above. After spontaneous 
breaking of symmetry in SM quarks acquire current masses through 
Higgs mechanism which entails additional contributions to hadron masses but 
parameters of gluon field between quarks remain the same. I.e., we suppose 
massless quarks to emit gluons in the same proportion as massive ones. As a 
result, parameters of gluon field in hadrons at present are the same as in 
chirally symmetric world. 
\subsection{Second scenario}
After spontaneous breaking of symmetry in SM massive quarks
emit gluons in other proportion than massless ones and parameters of gluon 
field in hadrons at present are different from those in chirally symmetric 
world. But for to evaluate the latter we should know, for example, pion mass 
in chirally symmetric world that is not equal to zero due to gluons as was 
discussed above.

\subsection{Concluding remarks}
At any rate, in either scenarios no additional mechanism of spontaneous 
symmetry breaking connected with Goldstone bosons is required. Another matter 
that massless quarks differ from each other only by their flavours and we 
come to the problem of origin for flavours. It is clear, however, the latter 
problem cannot be resolved within QCD which just takes flavours as given from 
outside. So problem of origin of flavours requires coming out from 
QCD-framework. By the way, one possible solution of this problem from cosmological 
positions was proposed by us a long time ago \cite{GonQCD}. 

To summarize, our confinement mechanism gives a physically reasonable approach 
to problem of chiral symmetry breaking without any additional mechanism of 
spontaneous symmetry breaking connected with Goldstone bosons. 

\section{A possible relation with a phenomenological string-like picture of 
confinement}
\subsection{The confining potential and string tension}
The results obtained in Sections 5--6 allow us to shed some light on one more 
problem which has been touched upon in Ref. \cite{Gon06}. 
As is known, during a long time up to now 
there exists the so-called string-like picture of quark
confinement but only at qualitative phenomenological level (see, e. g., 
Ref. \cite{Per}). Up to now, however, it is unknown as such a 
picture might be warranted from the point of view of QCD. Let us in short 
outline as our results (based on and derived from QCD-Lagrangian directly) 
naturally lead to possible justification of the mentioned construction. Thereto 
we note that one can calculate energy ${\cal E}$ of gluon condensate conforming 
to solution (3) in a volume $V$ through relation 
${\cal E}=\int_VT_{00}r^2\sin{\vartheta}dr d\vartheta d\varphi\>$ with $T_{00}$ 
of (25) but one should take into account that classical $T_{00}$ has a 
singularity along $z$-axis ($\vartheta=0,\pi$) and we have to introduce some 
angle $\vartheta_0$ so $\vartheta_0\leq\vartheta\leq\pi-\vartheta_0$.  
As well as in Ref. \cite{Gon051}, we may consider $\vartheta_0$ to be 
a parameter determining some cone $\vartheta=\vartheta_0$ so the quark  
emits gluons outside of the cone. Now if there are two quarks $Q_1, Q_2$ and 
each of them emits gluons outside 
of its own cone $\vartheta=\vartheta_{1,2}$ (see Figs. 1, 2) then we have 
soft gluons (as mentioned in Section 1) in regions I, II and between quarks.  

Accordingly, we shall have some region $V$ with gluon condensate between quarks 
$Q_1, Q_2$ and its vertical projection is shown in Fig. 1. Another projection 
of $V$ onto a plane perpendicular to the one of Fig. 1 is sketched out in 
Fig. 2.  
\begin{figure}
\vspace{0cm}
\caption{Vertical projection of region with the gluon condensate energy 
between quarks}
\end{figure}

\begin{figure}
\vspace{0cm}
\caption{Horizontal projection of region with the gluon condensate energy 
between quarks}
\end{figure}

Then, as is clear from Fig. 1, for distance $R$ between quarks we have 
$R=R_1\sin{\vartheta_1}+R_2\sin{\vartheta_2}$ and gluonic energy between
quarks will be equal to

$${\cal V}(R)=\int_VT_{00}r^2\sin{\vartheta}dr d\vartheta d\varphi=
\int_{r_1}^{R_1}\int^{\pi-\vartheta_1}_{\vartheta_1}
\int_{-\varphi_1}^{\varphi_1}
\left(\frac{{\cal A}}{r^2}+\frac{{\cal B}}{\sin^2{\vartheta}}\right)
\sin{\vartheta}dr d\vartheta d\varphi+$$
$$\int_{r_2}^{R_2}\int^{\pi-\vartheta_2}_{\vartheta_2}
\int_{-\varphi_2}^{\varphi_2}
\left(\frac{{\cal A}}{r^2}+\frac{{\cal B}}{\sin^2{\vartheta}}\right)
\sin{\vartheta}dr d\vartheta d\varphi
\>\eqno(35)$$
with constants ${\cal A}$, ${\cal B}$ defined in (25). 

To clarify a physical meaning of the quantities $r_{1,2}$ in Figs. 1, 2, 
let us recall an analogy with classical 
electrodynamics where is well known (see e. g. Ref. \cite{LL}) that the notion 
of classical electromagnetic field (a photon condensate) generated by a 
charged particle is applicable only at distances much greater than the Compton 
wavelength 
$\lambda_c=1/m$ for the given particle with mass $m$. Within the QCD framework 
the parameter $\Lambda_{QCD}$ plays a similar part (see, e.g., 
Ref. \cite{{pdg},{pi0}}). 
Namely, the notion of classical SU(3)-gluonic field ( a gluon condensate) is 
not applicable at the distances much 
less than $1/\Lambda_{QCD}$. In accordance with Section 3.1 we took 
$\Lambda_{QCD}=\Lambda=0.234$ GeV which entails $1/\Lambda\sim$ 0.8433 fm 
so one may consider 
$r_{1,2}\sim$ 0.1$r_0$ where $r_0$ is adduced in Tables 7--8 for 
charged pions and kaons. 

Under the circumstances, performing a simple integration in (35) with 
employing the relations $\int d\vartheta/\sin{\vartheta}=\ln\tan{\vartheta/2}$, 
$\tan{\vartheta/2}=\sin{\vartheta}/(1+ \cos{\vartheta})=
(1-\cos{\vartheta})/\sin{\vartheta}$, we shall without going into details 
(see also Ref. \cite{Gon051}) obtain  

$${\cal V}(R_1,R_2)={\cal V}_0-\sum_{i=1}^2
\frac{{4\varphi_i\cal A}\cos{\vartheta_i}}{R_i}+\sum_{i=1}^2 
{2\varphi_i\cal B}R_i\ln\frac{1+                 
\cos{\vartheta_i}}{1-\cos{\vartheta_i}}, \eqno(36)$$
where  ${\cal V}_0=\sum_{i=1}^2{\cal V}_{0i}=\sum_{i=1}^2
\left(\frac{{4\varphi_i\cal A}\cos{\vartheta_i}}{r_i}-
{2\varphi_i\cal B}r_i\ln\frac{1+\cos{\vartheta_i}}
{1-\cos{\vartheta_i}}\right)$. 

For the sake of simplicity let us put $R_1=R_2$, 
$\vartheta_1=\vartheta_2=\vartheta_0$, $\varphi_1=\varphi_2=\varphi_0$. Then 
$R_1=R_2=R/(2\sin{\vartheta_0})$ and from (36) it follows
$${\cal V}(R)= {\cal V}_0+\frac{a}{R}+kR \eqno(37)$$
with $a=-8\varphi_0{\cal A}\sin{2\vartheta_0}$, 
$k=2\varphi_0\frac{\cal B}{\sin{\vartheta_0}}
\ln{\frac{1+\cos{\vartheta_0}}{1-\cos{\vartheta_0}}}$. 

We recognize the modeling confining potential in (37) which is often used 
when applying to meson and heavy quarkonia physics (see, e.g., 
Refs. \cite{{Rob},{Bra}}). 
We can, however, see that phenomenological parameters 
$a, k, {\cal V}_0$ of potential (37) are expressed through more fundamental 
parameters $a_j$, $b_j$ connected with the unique exact solution (3) of 
Yang-Mills equations describing confinement. One can notice that the quantity 
$k$ (string tension) is usually related to the so-called Regge slope 
$\alpha^\prime=1/(2\pi k)$ and in many if not all of the papers using 
potential approach it is accepted $k\approx 0.18$ GeV$^2$ 
(see, e. g., Refs. \cite{{Rob},{Bra}}).  
\subsection{Estimates of $\vartheta_0$, $\varphi_0$ for charged pions and 
kaons}
Under the situation we have the equation
$$k=2\varphi_0\frac{\cal B}{\sin{\vartheta_0}}
\ln{\frac{1+\cos{\vartheta_0}}{1-\cos{\vartheta_0}}}\approx0.18 
\>\rm GeV^2\>\eqno(38)$$ 
with ${\cal B}=(b_1^2+b_1b_2+b_2^2)/2$, 
so let us employ (38) to estimate $\vartheta_0$, $\varphi_0$ if using the 
results obtained in Table 1 and 4 for charged pions and kaons and also for the 
ground state of toponium $\eta_t$ for that we use the 
parametrization from Ref. \cite{Gon08a} with the values 
$a_1= 0.361253$, $a_2= 0.339442$, $b_1= 48.9402$ GeV, $b_2= 76.7974$ GeV for 
the parameters of solution (3). Results of computations are presented in 
Table 9. 

\begin{table}[htbp]
\caption{Angular parameters determining the gluon condensate between quarks 
for charged pions, kaons and toponium ground state.}
\label{t.9}
\begin{center}
\begin{tabular}{|c|c|c|}
\hline
\small Particle 
& \small $\vartheta_0 $ 
& \small $\varphi_0$ \\ 
\hline
$\pi^{\pm}$---$u\overline{d}$, $\overline{u}d$ 
& \scriptsize $10^\circ$ 
& \scriptsize $14.76^\circ$ \\
\hline
& \scriptsize $30^\circ$ 
& \scriptsize $78.63^\circ$ \\
\hline
& \scriptsize $45^\circ$
& \scriptsize $166.16^\circ$ \\
\hline
& \scriptsize $60^\circ$
& \scriptsize $326.53^\circ$ \\
\hline
$K^{\pm}$---$u\overline{s}$, $\overline{u}s$ 
& \scriptsize $10^\circ$ 
& \scriptsize $7.76^\circ$ \\
\hline
& \scriptsize $30^\circ$ 
& \scriptsize $41.34^\circ$ \\
\hline
& \scriptsize $45^\circ$
& \scriptsize $87.36^\circ$ \\
\hline
& \scriptsize $60^\circ$
& \scriptsize $171.68^\circ$ \\
\hline
$\eta_t$---$\bar{t}t$ 
& \scriptsize $10^\circ$ 
& \scriptsize $(0.305\times10^{-4})^\circ$ \\
\hline
& \scriptsize $30^\circ$ 
& \scriptsize $(0.162\times10^{-3})^\circ$ \\
\hline
& \scriptsize $45^\circ$
& \scriptsize $(0.343\times10^{-3})^\circ$ \\  
\hline
& \scriptsize $60^\circ$
& \scriptsize $(0.675\times10^{-3})^\circ$ \\
\hline
& \scriptsize $80^\circ$
& \scriptsize $(0.240\times10^{-2})^\circ$ \\
\hline
& \scriptsize $88^\circ$
& \scriptsize $(0.123\times10^{-1})^\circ$ \\
\hline         
\end{tabular}
\end{center}
\end{table}        
If taking into account that only the values of $\vartheta_0$, $\varphi_0$ 
between 0 and $90^\circ$ are of physical meaning and, according to Figs. 1, 2, 
the corresponding region $V$ between quarks will be similar to a string-like 
one under the condition $\vartheta_0\to\pi/2$,  $\varphi_0\to0$, then we can 
see from Table 9 that the characteristic transverse sizes 
$D_{1,2}$ of the gluon condensate between quarks in fact tend to zero only 
in the case of heavy quarks, i.e., only for heavy quarks the gluon 
configuration between them might practically transform into a string. As a result, 
there arises the string-like picture of quark confinement but the latter seems 
to be warranted enough only for heavy quarks. It should be emphasized that 
string tension $k$ of (38) is determined just by parameters $b_{1,2}$ of linear 
magnetic colour field from solution (3) which indirectly confirms 
the dominant role of the mentioned field for confinement.

We cannot, however, speak about potential ${\cal V}(R)$ of (37) as describing 
some gluon configuration between quarks. 
It would be possible if the mentioned potential were a solution of Yang-Mills 
equations directly derived from QCD-Lagrangian since, from the QCD-point of 
view, any gluonic field should be a solution of Yang-Mills equations (as well 
as any electromagnetic field is by definition always a solution of Maxwell 
equations). 

In reality, as was shown in 
Refs. \cite{{Gon051},{Gon052}} (see also Appendix C in Ref. \cite{Gon08a}), 
potential of form (37) cannot be a solution 
of the Yang-Mills equations if simultaneously $a\ne0, k\ne0$. Therefore, 
it is impossible to obtain compatible solutions of the 
Yang-Mills-Dirac (Pauli, Schr{\"o}dinger) system when inserting potential of form 
(37) into Dirac (Pauli, Schr{\"o}dinger) equation. So, we draw the conclusion 
(mentioned as far back as in Refs. \cite{Gon03} and elaborated more in detail 
in Ref. \cite{Gon08a}) that the potential approach 
seems to be inconsistent: it is not based on compatible 
nonperturbative solutions for the Dirac-Yang-Mills system derived from 
QCD-Lagrangian in contrast to our confinement mechanism. Actually potential 
approach for heavy quarkonia has been historically modeled on positronium 
theory. In the latter case, however, one uses the {\em unique} modulo square 
integrable solutions of Dirac (Schr{\"o}dinger) 
equation in the Coulomb field [condensate of huge number of (virtual) photons], 
i. e., one employs the {\em unique} compatible nonperturbative solutions of the 
Maxwell-Dirac (Schr{\"o}dinger) system directly derived from QED-Lagrangian to 
describe positronium (or hydrogen atom) spectrum. 

To summarize, from the point of view of our approach both potential 
and string-like pictures of confinement arise only as some {\em effective} 
models derived in a certain way from the more fundamental theory based on 
exact solution (3) of SU(3)-Yang-Mills equations. This conclusion is in 
concordance with the preliminary one obtained in Ref. \cite{Gon06}. 

\section{Discussion and concluding remarks}
\subsection{Discussion}
 As is seen from Tables 7--8, at the characteristic scales
of charged pions and kaons the gluon concentrations are huge and the 
corresponding fields (electric and magnetic colour ones) can be considered to 
be the classical ones with enormous strengths. The part $n_{\rm coul}$ of gluon 
concentration $n$ connected with the Coulomb electric colour field is 
decreasing faster than $n_{\rm lin}$, the part of $n$ related to the linear 
magnetic colour field, and at large distances $n_{\rm lin}$ becomes dominant. 
It should be emphasized that in fact the gluon concentrations are much 
greater than the estimates given in Tables 7--8 because the latter are the 
estimates for maximal possible gluon frequencies, 
i.e. for maximal possible gluon impulses (under the concrete situation of 
charged pions and kaons). As was mentioned in Section 1, 
the overwhelming majority of gluons between quarks is soft, i. e., with 
frequencies much less than $\Gamma=1/\tau\approx$ $0.253\times10^{-7}$ eV, 
for example, in the case of pions, so the 
corresponding concentrations are much greater than those  
in Tables 7--8. The given picture is in concordance with the one obtained 
in Refs. \cite{{Gon03},{Gon08a},{Gon06},{Gon07a},{Gon07b}}. 
As a result, the confinement mechanism developed in 
Refs. \cite{{Gon01},{Gon051},{Gon052}} and described early in Section 1 is 
also confirmed by the considerations of the present paper. 

It should be noted, however, that our results are of a preliminary character 
which is readily apparent, for example, from that the current quark masses 
(as well as the gauge coupling constant $g$) used in computation are known only within the 
certain limits and we can expect similar limits for the magnitudes 
discussed in the paper so it is necessary further specification of the 
parameters for the confining SU(3)-gluonic field 
in charged pions and kaons which can be obtained, for instance, by calculating 
the width of decay $\pi^\pm\to\pi^0+e^\pm+\nu_e(\tilde{\nu}_e)$ 
with the help of wave function of $\pi^{0}$-meson discussed 
in Ref. \cite{Gon07a} and so on. We hope to continue analysing 
the given problems elsewhere. 

\subsection{Concluding remarks}
The results of present paper as well as the ones of Refs. 
\cite{{Gon03},{Gon08a},{Gon06},{Gon07a},{Gon07b}} allow one to speak about that the 
confinement mechanism elaborated in Refs. \cite{{Gon01},{Gon051},{Gon052}} 
gives new possibilities for considering many old problems of hadronic 
(meson) physics (such as nonperturbative computation of decay constants, masses 
and radii of mesons, chiral symmetry breaking and so forth) from the first 
principles of QCD immediately appealing 
to the quark and gluonic degrees of freedom. This is possible because the 
given confinement mechanism is based on the unique family of compatible 
nonperturbative solutions for the Dirac-Yang-Mills system directly derived from 
QCD-Lagrangian and, as a result, the approach is itself nonperturbative, 
relativistic from the outset, admits self-consistent nonrelativistic limit 
and may be employed for any meson (quarkonium). 

The given paper to a certain degree summarizes studying nonet of  
pseudoscalar mesons realized in Refs. \cite{{Gon06},{Gon07a},{Gon07b}} within 
the framework of our approach and we can ascertain the fact that, on the 
whole, this nonet can be described from the unified point of view of our 
confinement mechanism. In line with the above, obviously, one 
should now pass on to vector mesons ($\rho$, $\phi$, $\omega$...) and also to 
the light scalar mesons whose 
nature has been controversial over 30 years \cite{Close02}. As is clear from 
Section 2 and Appendices A, B, there exists a large number of relativistic 
bound states in the confining SU(3)-gluonic field (3) so all the mentioned 
mesons can probably correspond to some of those states and be described by 
their own sets of parameters $a_j$, $b_j$, $B_j$ 
of solution (3). Finally, one should think 
about possible ways to extend the approach over baryons, in particular, 
over nucleons.


\section*{Appendix A}
We here represent some results about eigenspinors of the Euclidean Dirac 
operator on two-sphere ${\Bbb S}^2$ employed in the main part of the paper. 

When separating variables in the Dirac equation (4) there naturally 
arises the Euclidean Dirac operator ${\cal D}_0$ on the unit two-dimensional 
sphere ${\Bbb S}^2$ and we should know its eigenvalues with the corresponding 
eigenspinors. Such a problem also arises in the black hole theory while 
describing the so-called twisted spinors on Schwarzschild and 
Reissner-Nordstr\"om black holes and it was analysed in 
Refs. \cite{{Gon052},{Gon99}}, so we can use the results obtained 
therein for our aims. Let us adduce the necessary relations. 

The eigenvalue equation for
corresponding spinors $\Phi$ may look as follows
$${\cal D}_0\Phi=\lambda\Phi.\>\eqno({\rm A}.1)$$

As was discussed in Refs. \cite{Gon99}, the natural form of ${\cal D}_0$ 
(arising within applications) in 
local coordinates $\vartheta, \varphi$ on the unit sphere ${\Bbb S}^2$ looks 
as 
$${\cal D}_0=-i\sigma_1\left[
i\sigma_2\partial_\vartheta+i\sigma_3\frac{1}{\sin{\vartheta}}
\left(\partial_\varphi-\frac{1}{2}\sigma_2\sigma_3\cos{\vartheta}
\right)\right]=$$
$$\sigma_1\sigma_2\partial_\vartheta+\frac{1}{\sin\vartheta}
\sigma_1\sigma_3\partial_\varphi- \frac{\cot\vartheta}{2}
\sigma_1\sigma_2         \eqno(\rm A.2)$$
with the ordinary Pauli matrices
$$\sigma_1=\pmatrix{0&1\cr 1&0\cr}\,,\sigma_2=\pmatrix{0&-i\cr i&0\cr}\,,
\sigma_3=\pmatrix{1&0\cr 0&-1\cr}\,, $$
so that $\sigma_1{\cal D}_0=-{\cal D}_0\sigma_1$.

The equation (A.1) was explored in Refs. \cite{Gon99}.
Spectrum of $D_0$ consists of the numbers
$\lambda=\pm(l+1)$              
with multiplicity $2(l+1)$ of each one, where $l=0,1,2,...$. Let us 
introduce the number $m$ such that $-l\le m\le l+1$ and the corresponding 
number $m'=m-1/2$ so $|m'|\le l+1/2$. Then the conforming eigenspinors of  
operator ${\cal D}_0$ are 
$$\Phi=\pmatrix{\Phi_1\cr\Phi_2\cr}= 
\Phi_{\mp\lambda}=\frac{C}{2}\pmatrix{P^k_{m'-1/2}\pm P^k_{m'1/2}\cr
P^k_{m'-1/2}\mp P^k_{m'1/2}\cr}e^{-im'\varphi}\> \eqno(\rm A.3) $$
with the coefficient $C=\sqrt{\frac{l+1}{2\pi}}$ and $k=l+1/2$.  
These spinors form an orthonormal basis in $L_2^2({\Bbb S}^2)$ 
and are subject 
to the normalization condition
$$\int_{{\Bbb S}^2}\Phi^{\dag}\Phi d\Omega=
\int\limits_0^\pi\,\int\limits_0^{2\pi}(|\Phi_{1}|^2+|\Phi_{2}|^2)
\sin\vartheta d\vartheta d\varphi=1\>. \eqno(\rm A.4)$$
Further, owing to the relation $\sigma_1{\cal D}_0=-{\cal D}_0\sigma_1$ we, 
obviously, have
$$ \sigma_1\Phi_{\mp\lambda}=\Phi_{\pm\lambda}\,.  \eqno(\rm A.5)$$

As to functions $P^k_{m'n'}(\cos\vartheta)\equiv P^k_{m',\,n'}(\cos\vartheta)$ 
then they can be chosen by 
miscellaneous ways, for instance, as follows (see, e. g.,
Ref. \cite{Vil91})
$$P^k_{m'n'}(\cos\vartheta)=i^{-m'-n'}
\sqrt{\frac{(k-m')!(k-n')!}{(k+m')!(k+n')!}}
\left(\frac{1+\cos{\vartheta}}{1-\cos{\vartheta}}\right)^{\frac{m'+n'}{2}}\,
\times$$
$$\times\sum\limits_{j={\rm{max}}(m',n')}^k
\frac{(k+j)!i^{2j}}{(k-j)!(j-m')!(j-n')!}
\left(\frac{1-\cos{\vartheta}}{2}\right)^j \eqno(\rm A.6)$$
with the orthogonality relation at $m',n'$ fixed
$$\int\limits_0^\pi\,{P^{*k}_{m'n'}}(\cos\vartheta)
P^{k'}_{m'n'}(\cos\vartheta)
\sin\vartheta d\vartheta={2\over2k+1}\delta_{kk'}
\>.\eqno(\rm A.7)$$
It should be noted that square of 
${\cal D}_0$ is 
$${\cal D}^2_0=-\Delta_{{\Bbb S}^2}I_2+
\sigma_2\sigma_3\frac{\cos{\vartheta}}{\sin^2{\vartheta}}\partial_\varphi
+\frac{1}{4\sin^2{\vartheta}} +\frac{1}{4}\>,
\eqno(\rm A.8)$$
while laplacian on the unit sphere is
$$\Delta_{{\Bbb S}^2}=
\frac{1}{\sin{\vartheta}}\partial_\vartheta\sin{\vartheta}\partial_\vartheta+
\frac{1}{\sin^2{\vartheta}}\partial^2_\varphi=
\partial^2_\vartheta+\cot{\vartheta}\partial_\vartheta
+\frac{1}{\sin^2{\vartheta}}\partial^2_\varphi\>,
\eqno(\rm A.9)$$
so the relation (A.8) is a particular case of the so-called 
Weitzenb{\"o}ck-Lichnerowicz formulas (see Refs. \cite{81}). 
Then from (A.1) it follows 
${\cal D}^2_0\Phi=\lambda^2\Phi$ and, when using the ansatz  
$\Phi=P(\vartheta)e^{-im'\varphi}=\pmatrix{P_1\cr P_2\cr}e^{-im'\varphi}$, 
$P_{1,2}=P_{1,2}(\vartheta)$, the equation ${\cal D}^2_0\Phi=\lambda^2\Phi$ 
turns into 
$$\left(-\partial^2_\vartheta-\cot{\vartheta}\partial_\vartheta +
\frac{m'^2+\frac{1}{4}}{\sin^2{\vartheta}}+
\frac{m'\cos{\vartheta}}{\sin^2{\vartheta}}\sigma_1\right)P=$$
$$\left(\lambda^2-\frac{1}{4}\right)P\>,
\eqno(\rm A.10)$$
wherefrom all the above results concerning spectrum of ${\cal D}_0$ can be 
derived \cite{Gon99}.

When calculating the functions $P^k_{m'n'}(\cos\vartheta)$ directly, to our 
mind, it is the most convenient to use the integral expression \cite{Vil91}

$$P^k_{m'n'}(\cos\vartheta)=\frac{1}{2\pi}
\sqrt{\frac{(k-m')!(k+m')!}{(k-n')!(k+n')!}}\>
\int_{0}^{2\pi}\left(e^{i\varphi/2}\cos{\frac{\vartheta}{2}}+
ie^{-i\varphi/2}\sin{\frac{\vartheta}{2}}\right)^{k-n'}\times$$
$$\left(ie^{i\varphi/2}\sin{\frac{\vartheta}{2}}+
e^{-i\varphi/2}\cos{\frac{\vartheta}{2}}\right)^{k+n'}e^{im'\varphi}d\varphi 
\eqno(\rm A.11)$$
and the symmetry relations ($z=\cos{\vartheta}$) 
$$P^k_{m'n'}(z)=P^k_{n'm'}(z), \>P^k_{m',-n'}(z)=P^k_{-m',\,n'}(z), 
\>P^k_{m'n'}(z)=P^k_{-m',-n'}(z)\,,$$ 
$$P^k_{m'n'}(-z)=i^{2k-2m'-2n'}P^k_{m',-n'}(z)\>. \eqno(\rm A.12)$$
In particular
$$P^{k}_{kk}(z)=
\cos^{2k}{(\vartheta/2)},  
P^{k}_{k,-k}(z)=i^{2k}\sin^{2k}{(\vartheta/2)},
P^{k}_{k0}(z)=\frac{i^{k}\sqrt{(2k)!}}{2^k k!}\sin^{k}{\vartheta}\,,$$
$$ P^{k}_{kn'}(z)=i^{k-n'}\sqrt{\frac{(2k)!}{(k-n')!(k+n')!}}
\sin^{k-n'}{(\vartheta/2)}\cos^{k+n'}{(\vartheta/2)}\>. \eqno(\rm A.13)$$ 
\subsection*{Eigenspinors with $\lambda=\pm1,\,\pm2$}
If $\lambda=\pm(l+1)=\pm1$ then $l=0$ and from (A.3) it follows that 
$k=l+1/2=1/2$, $|m'|\le1/2$ and we need the functions $P^{1/2}_{m',\pm1/2}$ 
that are easily evaluated with the help of (A.11)--(A.13) so   
the eigenspinors for $\lambda=-1$ are 
$$\Phi=\frac{C}{2}\pmatrix{\cos{\frac{\vartheta}{2}}+
i\sin{\frac{\vartheta}{2}}\cr
\cos{\frac{\vartheta}{2}}-i\sin{\frac{\vartheta}{2}}\cr}e^{i\varphi/2}, 
\Phi=\frac{C}{2}\pmatrix{\cos{\frac{\vartheta}{2}}+
i\sin{\frac{\vartheta}{2}}\cr
-\cos{\frac{\vartheta}{2}}+i\sin{\frac{\vartheta}{2}}\cr}
e^{-i\varphi/2},\eqno(\rm A.14)$$
while for $\lambda=1$ the conforming spinors are
$$\Phi=\frac{C}{2}\pmatrix{\cos{\frac{\vartheta}{2}}-
i\sin{\frac{\vartheta}{2}}\cr
\cos{\frac{\vartheta}{2}}+i\sin{\frac{\vartheta}{2}}\cr}e^{i\varphi/2}, 
\Phi=\frac{C}{2}\pmatrix{-\cos{\frac{\vartheta}{2}}+
i\sin{\frac{\vartheta}{2}}\cr
\cos{\frac{\vartheta}{2}}+i\sin{\frac{\vartheta}{2}}\cr}e^{-i\varphi/2}
\eqno(\rm A.15) $$
with the coefficient $C=\sqrt{1/(2\pi)}$.

It is clear that (A.14)--(A.15) can be rewritten in the form 
$$\lambda=-1: \Phi=\frac{C}{2}\pmatrix{e^{i\frac{\vartheta}{2}}
\cr e^{-i\frac{\vartheta}{2}}\cr}e^{i\varphi/2},\> {\rm or}\>\>
\Phi=\frac{C}{2}\pmatrix{e^{i\frac{\vartheta}{2}}\cr
-e^{-i\frac{\vartheta}{2}}\cr}e^{-i\varphi/2},$$
$$\lambda=1: \Phi=\frac{C}{2}\pmatrix{e^{-i\frac{\vartheta}{2}}\cr
e^{i\frac{\vartheta}{2}}\cr}e^{i\varphi/2}, \> {\rm or}\>\>
\Phi=\frac{C}{2}\pmatrix{-e^{-i\frac{\vartheta}{2}}\cr
e^{i\frac{\vartheta}{2}}\cr}e^{-i\varphi/2}\,, 
\eqno(\rm A.16) $$
so the relation (A.5) is easily verified at $\lambda=\pm1$. 

In studying vector mesons and excited states of heavy quarkonia eigenspinors 
with $\lambda=\pm2$ may also be useful. Then $k=l+1/2=3/2$, $|m'|\le3/2$ and we 
need the functions $P^{3/2}_{m',\pm1/2}$ 
that can be evaluated with the help of (A.11)--(A.13). Computation gives 
rise to 
$$ P^{3/2}_{3/2,-1/2}=-\frac{\sqrt{3}}{2}\sin{\vartheta}
\sin{\frac{\vartheta}{2}}= P^{3/2}_{-3/2,1/2},\>$$   
$$P^{3/2}_{3/2,1/2}=i\frac{\sqrt{3}}{2}\sin{\vartheta}
\cos{\frac{\vartheta}{2}}= P^{3/2}_{-3/2,-1/2},\>$$
$$P^{3/2}_{1/2,-1/2}= -\frac{i}{4}\left(\sin{\frac{\vartheta}{2}}-
3\sin{\frac{3}{2}\vartheta}\right)=P^{3/2}_{-1/2,1/2},\>$$
$$P^{3/2}_{1/2,1/2}= \frac{1}{4}\left(\cos{\frac{\vartheta}{2}}+
3\cos{\frac{3}{2}\vartheta}\right)=P^{3/2}_{-1/2,-1/2},\>
\eqno(\rm A.17) $$
and according to (A.3) this entails eigenspinors with $\lambda=2$ in the 
form
$$\frac{C}{2}i\frac{\sqrt{3}}{2}\sin{\vartheta}
\pmatrix{e^{-i\frac{\vartheta}{2}}\cr
e^{i\frac{\vartheta}{2}}\cr}e^{i3\varphi/2},\>
\frac{C}{8}\pmatrix{3e^{-i\frac{3\vartheta}{2}}+e^{i\frac{\vartheta}{2}}\cr
3e^{i\frac{3\vartheta}{2}}+e^{-i\frac{\vartheta}{2}}\cr}e^{i\varphi/2},\>$$
$$\frac{C}{8}\pmatrix{-3e^{-i\frac{3\vartheta}{2}}-e^{i\frac{\vartheta}{2}}\cr
3e^{i\frac{3\vartheta}{2}}+e^{-i\frac{\vartheta}{2}}\cr}e^{-i\varphi/2},\>
\frac{C}{2}i\frac{\sqrt{3}}{2}\sin{\vartheta}
\pmatrix{-e^{-i\frac{\vartheta}{2}}\cr
e^{i\frac{\vartheta}{2}}\cr}e^{-i3\varphi/2}\>
 \eqno(\rm A.18) $$
with $C=1/\sqrt{\pi}$, while eigenspinors with $\lambda=-2$ are obtained in 
accordance with relation (A.5). 

\section*{Appendix B}
We here adduce the explicit form for the radial parts of meson wave functions 
from (6). At $n_j=0$ they are given by 
$$F_{j1}=C_jP_jr^{\alpha_j}e^{-\beta_jr}\left(1-
\frac{Y_j}{Z_j}\right),F_{j2}=iC_jQ_jr^{\alpha_j}e^{-\beta_jr}\left(1+
\frac{Y_j}{Z_j}\right),\eqno(\rm B.1)$$
while at $n_j>0$ by
$$F_{j1}=C_jP_jr^{\alpha_j}e^{-\beta_jr}\left[\left(1-
\frac{Y_j}{Z_j}\right)L^{2\alpha_j}_{n_j}(r_j)+
\frac{P_jQ_j}{Z_j}r_jL^{2\alpha_j+1}_{n_j-1}(r_j)\right],$$
$$F_{j2}=iC_jQ_jr^{\alpha_j}e^{-\beta_jr}\left[\left(1+
\frac{Y_j}{Z_j}\right)L^{2\alpha_j}_{n_j}(r_j)-
\frac{P_jQ_j}{Z_j}r_jL^{2\alpha_j+1}_{n_j-1}(r_j)\right]\eqno(\rm B.2)$$
with the Laguerre polynomials $L^\rho_{n}(r_j)$, $r_j=2\beta_jr$, 
$\beta_j=\sqrt{\mu_0^2-\omega_j^2+g^2b_j^2}$ at $j=1,2,3$ with 
$b_3=-(b_1+b_2)$, 
$P_j=gb_j+\beta_j$, $Q_j=\mu_0-\omega_j$,
$Y_j=P_jQ_j\alpha_j+(P^2_j-Q^2_j)ga_j/2$, 
$Z_j=P_jQ_j\Lambda_j+(P^2_j+Q^2_j)ga_j/2$    
with $a_3=-(a_1+a_2)$,   
$\Lambda_j=\lambda_j-gB_j$ with $B_3=-(B_1+B_2)$, 
$\alpha_j=\sqrt{\Lambda_j^2-g^2a_j^2}$, 
while $\lambda_j=\pm(l_j+1)$ are
the eigenvalues of Euclidean Dirac operator ${\cal D}_0$ 
on unit two-sphere with $l_j=0,1,2,...$ (see Appendix A) 
and quantum numbers $n_j=0,1,2,...$ are defined by the relations 
$$n_j=\frac{gb_jZ_j-\beta_jY_j}{\beta_jP_jQ_j}\,, 
\eqno(\rm B.3)$$
which entails the quadratic equation (7) and spectrum (8).  
Further, $C_j$ of (B.1)--(B.2) should be determined
from the normalization condition
$$\int_0^\infty(|F_{j1}|^2+|F_{j2}|^2)dr=\frac{1}{3}\>.\eqno(\rm B.4)$$
As a consequence, we shall gain that in (6) 
$\Psi_j\in L_2^{4}({\Bbb R}^3)$ at any $t\in{\Bbb R}$ and, accordingly,
$\Psi=(\Psi_1,\Psi_2,\Psi_3)$ may describe relativistic bound states 
in the field (3) with the energy spectrum (8). As is clear from (B.3) at 
$n_j=0$ we have 
$gb_j/\beta_j=Y_j/Z_j$ so the radial parts of (B.1) can be rewritten as  
$$F_{j1}=C_jP_jr^{\alpha_j}e^{-\beta_jr}\left(1-
\frac{gb_j}{\beta_j}\right),F_{j2}=iC_jQ_jr^{\alpha_j}e^{-\beta_jr}\left(1+
\frac{gb_j}{\beta_j}\right)\>.\eqno(\rm B.5)$$
More details can be found in Refs. \cite{{Gon01},{Gon052}}. 


\end{document}